\RequirePackage{fix-cm}
\documentclass[smallextended]{svjour3}       
\smartqed  
\usepackage{graphicx}
\usepackage[breaklinks=True]{hyperref}
\begin{document}

\title{Monitoring the photometric behavior of OmegaCAM with Astro-WISE}

\titlerunning{OmegaCAM and Astro-WISE}        

\author{Verdoes Kleijn G.A., Kuijken K.H., Valentijn E.A., Boxhoorn D.R., Begeman K.G., Deul E.D., Helmich E.M., \& Rengelink R. }

\authorrunning{Verdoes Kleijn et al.} 

\institute{G. Verdoes Kleijn, E. Valentijn, D. Boxhoorn, K. Begeman, R. Rengelink \at
              Kapteyn Astronomical Institute, University of Groningen, P.O. Box 800, 9700 AV Groningen, The Netherlands
           \and
           K. Kuijken, E. Deul \at
              Leiden Observatory, Leiden University, P.O. Box 9513, 2300 RA Leiden, The Netherlands
}

\date{Received: date / Accepted: date}

\maketitle

\begin{abstract}
The OmegaCAM wide-field optical imager is the sole instrument on the VLT Survey Telescope at ESO's Paranal Observatory. The instrument, as well as the telescope, have been designed for surveys with very good, natural seeing-limited image quality over a 1 square degree field. OmegaCAM was commissioned in 2011 and has been observing three ESO Public Surveys in parallel since October 15, 2011. We use the Astro-WISE information system to monitor the calibration of the observatory and to produce the Kilo Degree Survey (KiDS).  

Here we describe the photometric monitoring procedures in Astro-WISE and give a first impression of OmegaCAM's photometric behavior as a function of time. The long-term monitoring of the observatory goes hand in hand with the KiDS survey production in Astro-WISE. KiDS is observed under partially non-photometric conditions. Based on the first year of OmegaCAM operations it is expected that a $\sim 1\%-2\%$ photometric homogeneity will be achieved for KiDS. 

\keywords{imager, wide-field system \and survey system \and VLT/VST \and Astro-WISE \and information system}
\end{abstract}



\section{Public Surveys of the Southern Hemisphere}

The European Southern Observatory is in the process of delivering to its community three Public Surveys that image the Southern Hemisphere: the Kilo Degree Survey~\cite{deJong12}, ATLAS and VPHAS+ \cite{arnaboldi07}. The surveys are performed using the OmegaCAM wide-field optical imager with 1 square degree FoV. It is the sole instrument on the 2.6m VLT Survey Telescope at Paranal Observatory, Chile. 

The three ESO Public Surveys span thousands of square degrees of sky obtained over years of observations. For example, the Kilo Degree Survey will map 1500 square degrees of extragalactic sky in the Sloan u, g, r and i passbands. This requires $\sim 440$ "Full Night Equivalents" spread over at least 3 years. The scientific drivers for the survey ask for $\sim 1\%$ accuracy on the photometric scale, in absolute sense and between passbands (see \cite{deJong12} in this issue). This means that all systematic variations along the signal path (atmosphere, telescope, instrument) must be characterized to a fraction of that $1\%$, consistently over the many years of survey operation. This includes for example Paranal's atmospheric variability, observed to have peak-to-peak variations of up to $\sim 0.07$mag/airmass under clear sky conditions \cite{patat11}; systematic changes in mirror reflectivity (dusting), with typical rates of few tenths of magnitude per year and subtle variations in the system's electronic gain.

OmegaCAM and VST have been commissioned in 2011 \cite{kuijken11}. Survey operations started October 15, 2011 with the three surveys being performed in parallel. Operations are based on a strict observing protocol for both calibrations and surveys. We use the Astro-WISE information system to monitor the calibration of the observatory and to produce and deliver the Kilo Degree Survey. This paper presents the approach implemented in Astro-WISE for photometric monitoring OmegaCAM to maximize the quality of its survey imaging products. In Astro-WISE, the homogeneous stream of data flowing from the instrument's Calibration Plan is turned into  pixeldata and metadata through which the full physical system of Paranal's atmosphere + VST + OmegaCAM is continuously monitored. The design of Astro-WISE has been tuned explicitly to have survey production take advantage dynamically and very much automatically of increases in calibration accuracy and improved insight of the physical system from long-term trend analyses. In other words, the information loop between calibrating an instrument and calibrating survey data is closed inside the Astro-WISE system. 

The paper is ordered as follows. We start out by describing the OmegaCAM instrument in its commissioned form in Section~\ref{s:omegacamAtVST}. We give an overview of OmegaCAM's dataflow (Section~\ref{s:dataflow}). We then present the Photometric Calibration Plan (Section~\ref{s:calibrationPlan}) and the conceptual approach in Astro-WISE for photometric monitoring based on this plan (Section~\ref{s:monitoringInAstroWISE}). This is combined with preliminary results of OmegaCAM's photometric behavior in the first $\sim$6 months of operations. We end with conclusions and outlook (Section~\ref{s:conclusionsAndOutlook}). 

\section{OmegaCAM at the VST}
\label{s:omegacamAtVST}

The VLT Survey Telescope and OmegaCAM have both been designed specifically to have (i) the stability to produce homogeneous surveys with long-term observational programs and (ii) image quality as one of the major scientific strengths. 

\begin{figure}[ht]
\includegraphics[width=10.0cm]{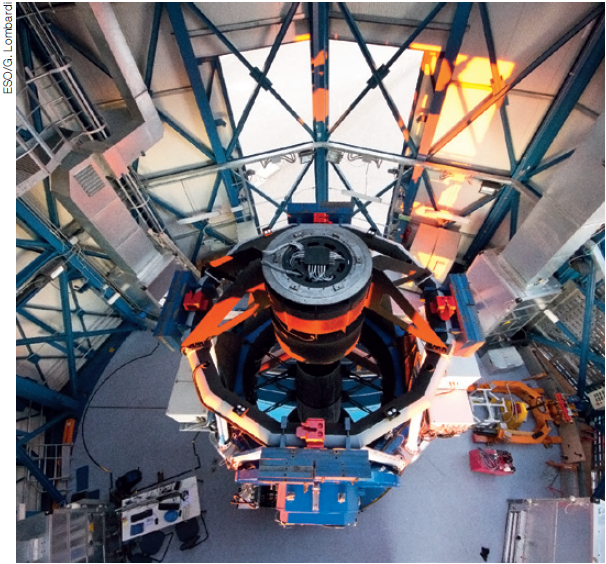}
\caption{The VST from  above. The M1 cover is open and the primary mirror is visible with its baffle. On top, the back side of the hexapod driving M2 is visible. The four sets of two domelamps are also visible, mounted on the white trusses forming a dodecagon.}
\label{f:vstFromAbove}
\end{figure}

The VLT Survey Telescope \cite{capaccioli11} is of modified Ritchey-Cretien design and has an alt-az mount . The VST telescope can work in two configurations. In the standard configuration a two-lens field corrector is used. For very large zenith distances the second configuration replaces this corrector with one including an Atmospheric Dispersion Corrector (ADC), consisting of one lens and two counter-rotating prism pairs. The operating wavelength ranges are 320-1014nm and 365-1014nm for the two-lens corrector and corrector + ADC respectively. The VST has active primary and secondary mirrors (Figure~\ref{f:vstFromAbove}). 

\begin{figure}[ht]
\includegraphics[width=10.0cm]{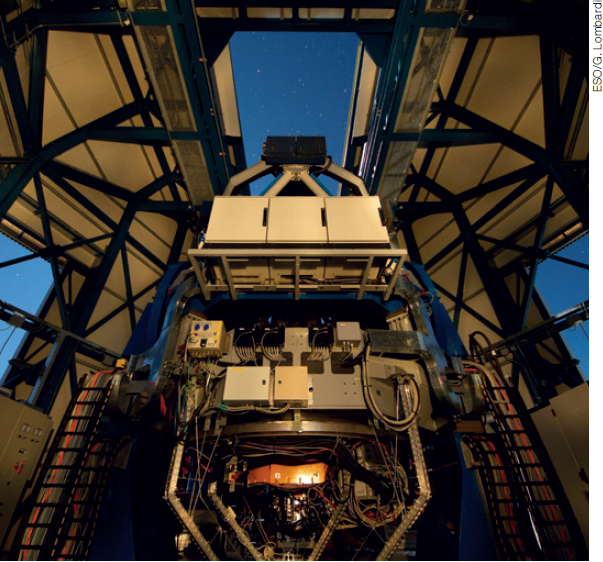}
\caption{The VST as seen from the observing floor. The OmegaCAM instrument and its control cabinets are in the lower section, interfaced to the telescope at the Cassegrain focus.}
\label{f:vstFromBelow}
\end{figure}

\begin{figure}[ht]
\includegraphics[width=10.0cm]{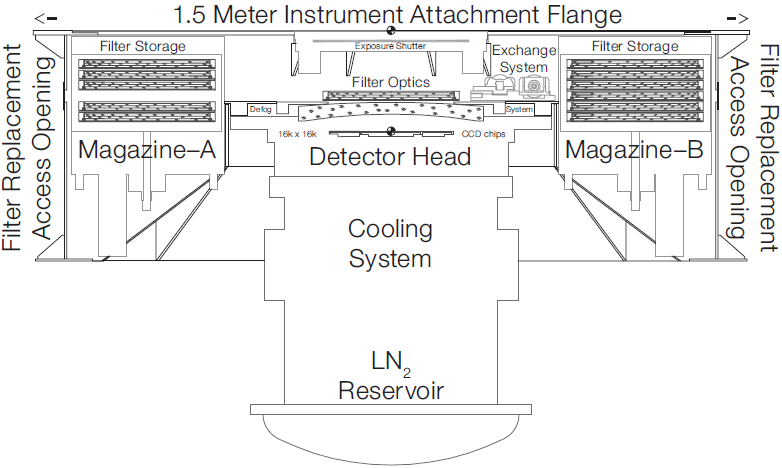}
\caption{Schematic overview of the main components of OmegaCAM. The 1.5m diameter housing structure contains various labeled components. From top to bottom: the shutter followed by the filter housing. The filters are supplied by two magazines. Each magazine can be filled with maximally 6 filters. The large, curved dewar window (the final optical element in the VST design) is followed by the detectors. The dewar is cooled using a 40-l Nitrogen cryostat.} 
\label{f:omegacamSchematicOverview}
\end{figure}

OmegaCAM is mounted at the Cassegrain focus (Figure~\ref{f:vstFromBelow}). The optical parts located in the instrument are the filters, and the dewar window to the cryostat, which doubles as a field lens (Figure~\ref{f:omegacamSchematicOverview}). In front of the dewar window is the mechanical part of OmegaCAM: closest to the window sits the
filter exchange mechanism, and above that the shutter. The housing provides the mechanical link between the telescope and the detector/cryostat system. 
OmegaCAM is a 268 Megapixel camera with a 1 square degree FoV.  The optical design of the camera has little aberration over the full field and a constant 0.21'' plate scale. Wavefront sensing is done in OmegaCAM. The wavefront sensors work by registering star images that are significantly out of focus: the two auxiliary CCDs are mounted out of the focal plane, one 2 mm above and one 2 mm below. The resulting PSF is observed to equal the atmospheric seeing down to 0.6'' over the full FoV. Paranal offers such excellent seeing much of the time (a 50 \% fractile of 0.66'' FWHM\footnote{http://www.eso.org/sci/facilities/paranal/site/paranal.html}). 

\subsection{Calibration unit}
The VST enclosure houses an in-dome photometric calibration unit. It consists of two sets of 4 lamps each. They are commercial 12-24V halogen lamps mounted on the trusses on which the secondary mounting rests (Figure~\ref{f:vstFromAbove}). The lamps illuminate a dome screen, that is located on the inside of the enclosure near zenith. Each set is operated independently and is using a stabilized current supply, one unit per set. Lamps are switched on/off with a
gradual increase/decrease of the current over a timespan of 3-5 minutes. Lamp intensity variations are
observed to be smaller than 1\% over a month (see Section~\ref{s:monitoringInAstroWISE}).
The two sets are operated at different rates, one set $\sim 0.75$ hour/day the
other set at $\sim 0.75$ hour/week. This ensures continuous calibration coverage. The calibration unit serves as an internal standard candle for a range of instrument monitoring and calibration activities. Dome screen observations are used to perform a daily overall "Quick Health Check", to measure pixel-to-pixel sensitivity variations, to measure CCD linearity and to check shutter timing accuracy.

\subsection{Shutter unit}
\begin{figure}[ht]
\includegraphics[width=10.0cm]{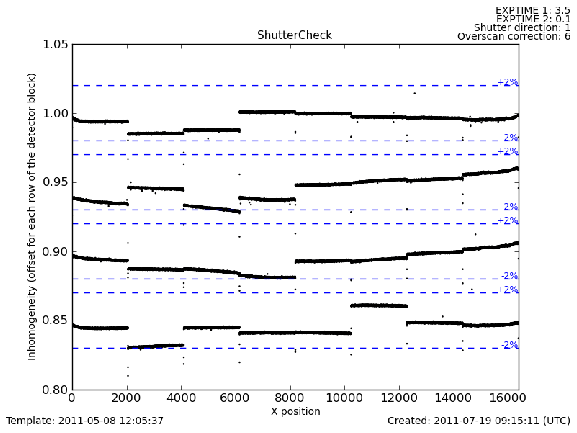}
\caption{Ratio image of flux in a 0.1 sec domeflat exposure in Sloan z divided by the flux in a 3.5 sec exposure.
The shutter blades move along the x-axis in the same direction for both exposures. For each CCD the average
flux per column is computed. The results for the top row of 8 detectors is plotted at its right location. The
subsequent rows below it are offset by 0.05. The two dashed blue lines for each row indicate $\pm2\%$ deviation. The systematics in the plots, discrete jumps from one CCD to another, are dominated by the non-linearity of the detectors. 
}
\label{f:shutterCheck}
\end{figure}

The exposure shutter covers an aperture size of 370$\times$292mm. It consists of two carbon fiber blades. They are driven by 2 micro-stepper motors and move smoothly. These movements are controlled such that each individual CCD pixel `sees' the opening edge of the one blade and the closing edge of the other blade with an identical time difference, even if the blades are still accelerating. This provides an impact-free, high-accuracy photometric shutter. Commissioning tests have shown that for an exposure time of 0.2sec / 1sec, deviations from a homogeneous exposure are below 2\% / 0.2\% over the whole field of view. This is based on observed variation in the flux level ratio between domeflat exposures as a function of position in the detector mosaic (Figure~\ref{f:shutterCheck}). The domeflats are taken under identical conditions, but with short and long exposure times. This gives an upper limit to photometric effect of shutter motion inhomogeneity as several other sources contribute to variations in flux level ratio (e.g., calibration unit photometric variations, non-linear behavior of detectors).

\subsection{Filter unit}
\begin{figure}[ht]
\includegraphics[width=10.0cm]{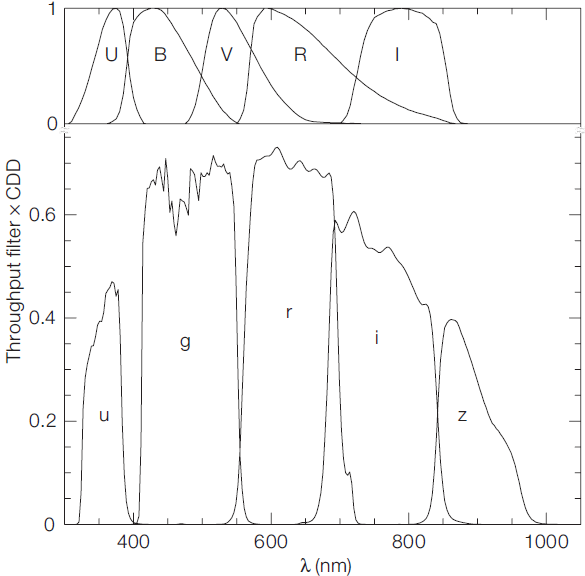}
\caption{The throughput of the entire VST system for the broadband Sloan filters in OmegaCAM is shown (lower) with the passbands of the standard Johnson-Cousins UBVRI filters shown for comparison (upper).}
\label{f:omegacamTransmissionCurves}
\end{figure}

The filters are stored in two magazines which can move up and down on either side of the focal plane, through large shafts in the housing (Figure~\ref{f:omegacamSchematicOverview}). A linear stage slides filters into the beam, where they are locked into place by means of movable notches. The total number of motors used in the filter exchange and positioning mechanism is seven. Commissioning tests verify that the precision of filter positioning is good. Variations in photometry due to optical imperfections in the filters are less than 0.3\% over $\sim 5$ months of operations. This has been verified for the u, g, r and i filter. The conclusion is based on inspection of ratios in flux level of domeflats observed over timescales of months. The long-term monitoring of domeflats is presented in Section~\ref{s:domeflatfielding} and results are shown also in Figure~\ref{f:domeflat}. To minimize operational overhead time the filter exchange unit is built in such a way that it allows one filter to be pulled into the beam while the previous one is pushed out. The filters are large and heavy: when fully loaded with 12 filters, the instrument contains 40 kg of filter glass alone! During the filter exchange process about 17kg are moved (mass of 2 filters and the carriage).
The primary filter set of OmegaCAM are Sloan u', g', r', i' and z' filters. In addition, there are Johnson B and V filters for stellar work and for cross-calibrating the photometric systems, a Stromgren v filter, an H$\alpha$ filter consisting of 4 segments with redshifts of up to 10000km/sec, and a segmented ugri filter for efficient photometric monitoring of the sky. Filters have been manufactured by SAGEM in Paris, and by Barr Associates in Massachusetts, and consist of 3-layer sandwiches of coloured or coated glass plates. The Sloan filters are interference filters. Figure~\ref{f:omegacamTransmissionCurves} shows transmission curves for the Sloan filters. 

\subsection{CCD detectors}
\begin{figure}[ht]
\includegraphics[width=10.0cm]{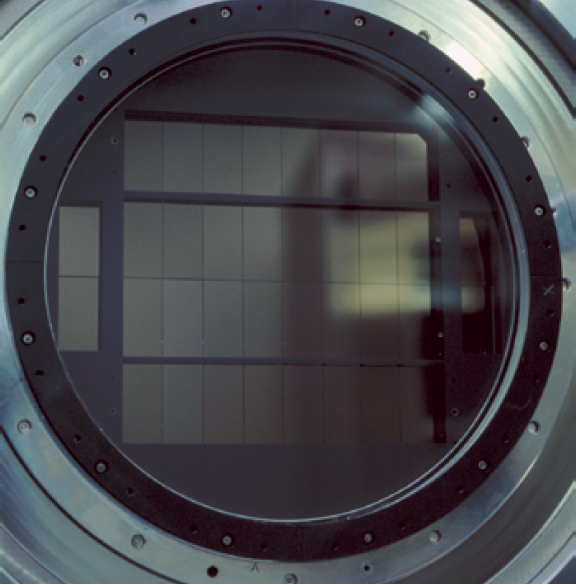}
\caption{The detector mosaic at the heart of OmegaCAM in its dewar ready for installation in the instrument. In addition to the 'science array' of 32 CCDs, the 4 'auxiliary' CCDs for guiding and image quality analysis are seen.}
\label{f:omegacamFrontView}
\end{figure}

\begin{figure}[ht]
\includegraphics[width=10.0cm]{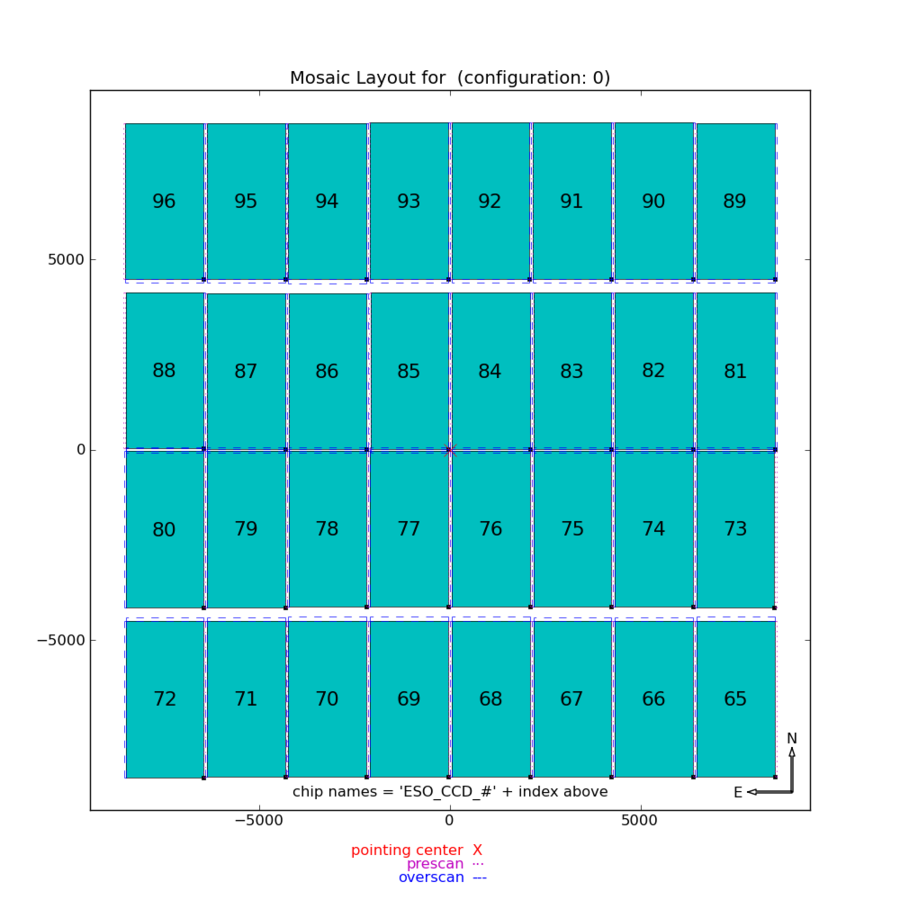}
\caption{Schematic outline of the 32 CCDs in the OmegaCAM detector mosaic used for science observations. The chip numbering as used by Astro-WISE and ESO is indicated together with the on-sky orientation. Each detector is read out with a pre- and over-scan area in X and a overscan area in Y.}
\label{f:layoutOmegaCAMdetectors}
\end{figure}

The OmegaCAM optical CCD imaging camera consists of a 'science array' of 32 2k x 4k E2V 44-82 devices for a total area of 16384 x 16384 15$\mu$ pixels, corresponding to 26x26cm (Figure~\ref{f:layoutOmegaCAMdetectors}). The CCDs are thinned and low-noise (5e${-}$ read-out noise). They are 3-edge buttable and cover the 1x1 square degree field of view of the VST with small gaps (25''-85'') and have a pixelsize of 0.21 arcsec/pixel. Around this science array lie four `auxiliary CCDs', of the same format. The mosaic has been built at ESO's headquarters in Garching.
The whole detector system is shown in Figure~\ref{f:omegacamFrontView}. Two of the auxiliary CCDs, on opposite sides of the science array, are used for auto-guiding using a guide star. The field is so large that also field rotation is auto-guided. The other two CCDs are used for wavefront sensing using real-time image analysis. 

\section{OmegaCAM Data flow from Paranal to Astro-WISE}
\label{s:dataflow}

The daily output of OmegaCAM is currently $\sim 70$Gbyte of raw imaging data on average, corresponding to 140 exposures. This includes day and night-time calibration and science exposures. This data stream of over 2 Terabyte per month is automatically pulled into the Astro-WISE data handling system, situated approximately 11250 kilometers from the imager. OmegaCAM data is transferred by ESO from Paranal Observatory, Chile to headquarters in Garching, Germany using EVALSO \cite{filippi10}. From there data is pulled automatically to the Astro-WISE node in Groningen via monitoring (using wget in mirror mode). The nature of the data (bias, science, twilight, etc...) is determined from header inspection of FITS format data in Groningen. This also intercepts data not to be ingested in Astro-WISE (e.g., focus runs). Pixel data is then decompressed. The new raw data triggers ingestion into Astro-WISE: pixel data is stored on dataservers and metadata (i.e., non-pixel data) in the database (see \cite{verdoesKleijn11} for details) and an initial rough quality assessment is automatically performed. Via this retrieval approach we have succeeded in having OmegaCAM data available for processing in Astro-WISE as fast as 1 hour after its observation. A week's worth of calibration data is processed and quality assessed by a single calibration scientist in about 1 day. For the KiDS science data, a Quick Look pipeline monitors for new raw data and automatically produces coadded calibrated images. Latest instead of default calibrations are automatically used if available. At the Astro-WISE node in Groningen we have achieved production of up to 70 square degrees of coadded calibrated science images in 24 hours.

Swift transport and ingestion of new data in Astro-WISE is important for instrument monitoring and survey efficiency. Every week of delay in noticing degradation of nightly data quality and/or instrument performance corresponds to a loss of $\sim 2\%$ for a survey per year. The science yield and rate is also enlarged by fast production and analysis of end-products. For the OmegaCAM KiDS survey for example, one can think of swift follow-up observations (instead of 1 year waits for return of the same RA range) of highly competitive fields such as high redshifts QSOs and extremely cool brown dwarfs. 

\section{OmegaCAM Photometric Calibration Plan}
\label{s:calibrationPlan}

\begin{figure}[ht]
\includegraphics[width=10.0cm]{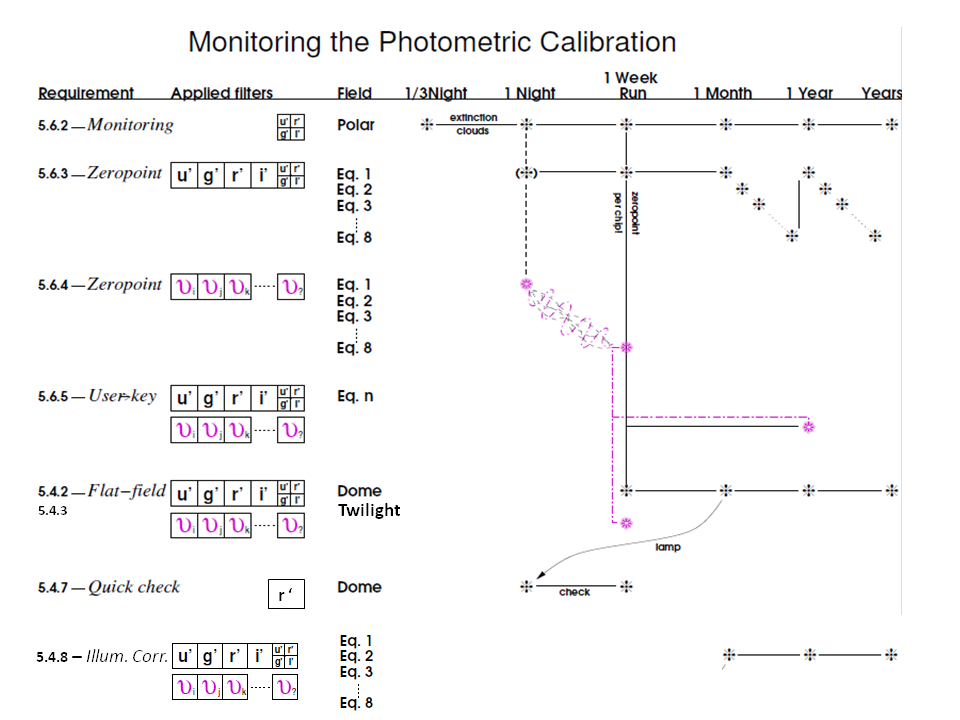}
\caption{Graphical table overview how the photometric calibration of both OmegaCAM+VST and the atmosphere is monitored. From left
to right, the table indicates the requirement number and name as used in Astro-WISE an the Calibration Plan, the used filters ($\nu_i$ indicating user bands), the used fields (SA and polar fields). The stars indicate at which frequency the measurement is done. For further details, see text.}
\label{f:calibrationPlan}
\end{figure}

The Photometric Calibration Plan of OmegaCAM\footnote{See http://www.eso.org/sci/facilities/paranal/instruments/omegacam/doc for Calibration Plan and User Requirements.} is designed to provide a continuous photometric characterization of the survey system. A schematic overview of OmegaCAM's photometric Calibration Plan is shown in Figure~\ref{f:calibrationPlan}. The types of observations allow to establish the photometric scale of the OmegaCAM+VST system and atmosphere separately and the observational cadences should allow continuous monitoring, i.e., sample the expected timescales of variations. The system is monitored using the calibration unit with a daily ({\it Quick check}) and weekly ({\it Dome Flat-Field}) frequency. These observations are the input to determine the stability and systematic photometric behavior of OmegaCAM+VST system. On-sky observations monitor the system plus atmosphere, using fixed fields with frequencies of three-per-night ({\it Monitoring}), nightly/weekly ({\it Zeropoint},{\it Twilight Flatfield}) and monthly or longer ({\it User$\rightarrow$Key}, {\it Illum.~Corr.}). Thus the photometric behavior of the atmosphere as a function of time can be deduced from the combination of on-sky and in-dome observations. 

\section{Monitoring OmegaCAM with Astro-WISE}
\label{s:monitoringInAstroWISE}

\begin{figure}[ht]
\includegraphics[width=10.0cm]{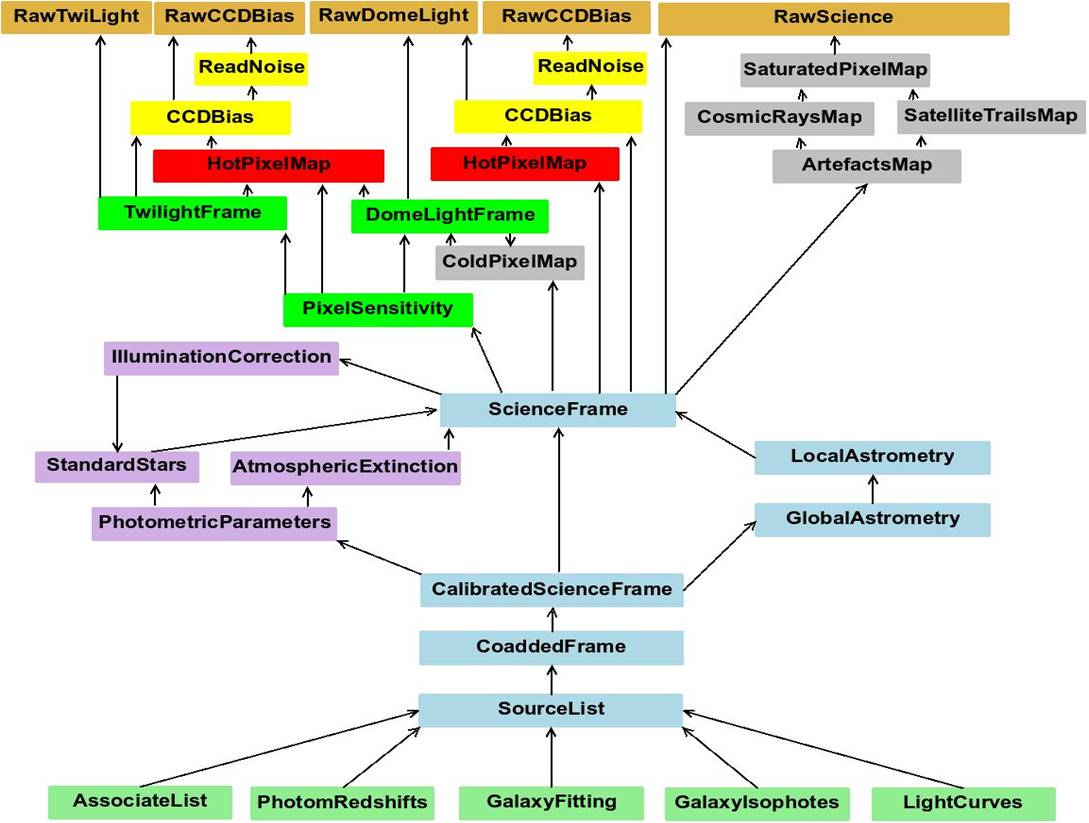}
\caption{Overview of classes of data objects. Data objects not only
  contain the survey products denoted by familiar names in wide-field
  imaging. They also carry the information how they, as requested
  target, can be created out of other survey objects, illustrated by
  the arrows. Underlying is an object model that captures the
  relationship between requested information and the physics of the
  atmosphere-to-detector observational system.}
\label{f:targetDiagram}
\end{figure}

We use the Astro-WISE information system to handle the observations from the OmegaCAM Photometric Calibration Plan. A major design goal of Astro-WISE has been to have Public Survey production take advantage dynamically and very much automatically of improved calibration accuracy and improved insight of the physical system inferred from long-term monitoring using observations from the Photometric Calibration Plan. In other words, we have closed the information loop between calibrating an instrument and calibrating a survey inside a single system of information. We refer the reader to \cite{begeman12} (this volume) and \cite{valentijn07} for a comprehensive description of this Astro-WISE design. The approach taken in Astro-WISE is to map the calibrational requirements (to be achieved by the Calibration Plan) to a data model. The data model is a map that describes for each calibration requirement which information it needs as input to be tested. This information is expressed as products from types of observations of the Calibration Plan. In other words, the data model maps which type of observations are needed to construct the desired information. This mapping is a cascade from final products back to the initial raw observations taken for the Calibration Plan (see Figure~\ref{f:targetDiagram}). The data model is conceptual: it does not fix the content of a method, e.g., the algorithm, to make each step in this cascade.

This conceptual data model is implemented as an object model. This means that each photometric requirement has a corresponding class or classes (e.g., 
{\it Zeropoints} $\rightarrow$ {\tt PhotometricParameters}, 
{\it Monitoring} $\rightarrow$ {\tt AtmosphericExtinctionCoefficients}). 
The classes aggregate the information required to test the requirement. Objects are specific instantiations of classes. For example, each set of exposures that represent a component of the Calibration Plan (i.e., Observing Blocks in ESO speak) is represented as one object. Objects can contain both meta-data and (processed) pixel data. Meta-data is defined as all non-pixel data: e.g., observational metadata, processing configurations, image statistics and quality assessment parameters.  The objects have links to their dependencies recursively, where dependencies are other objects on which it depends for information input. Thus, objects representing final survey products can be traced back to their set of raw data dependencies from which they are ultimately derived. This is the concept that eventually links requirements on calibrational accuracy in the Calibration Plan to attributes of objects stored in the Astro-WISE database. Trend plots of these attributes as function of time, any other attribute of the object, or any attribute of its dependencies is the basis of quality assessment and control. The Python Language is used to implement the object model\footnote{The Astro-WISE Python documentation server gives at http://doc.astro-wise.org/astro.main.html an IT-technical description of the Astro-WISE object model.}

In essence, Astro-WISE is an integrated information accumulator for survey data handling. Not a pipeline feeding a separate archive feeding a separate set of data analysis tools. It keeps the cycle processing$\rightarrow$ archiving $\rightarrow$ in-depth monitoring analysis around the pool of data (observational pixels and metadata) in a single system. Analysis by survey team members and calibration scientists on the data results in an improved understanding of the instrument and how it should be calibrated. Changing methods is one way in which this improved knowledge can be captured by Astro-WISE. This does not alter the object model.  By applying the improved knowledge (i.e., reprocessing) the next cycle starts to increase the level of detail to which the instrument plus the processing of its information is understood.

The next sections describe in more detail the observations for each component in the Calibration Plan and how they are handled in Astro-WISE. Preliminary monitoring results are presented. They are based on the first $\sim$ 6 months of operations.

\subsection{Quick check}
\label{s:inDome}

\begin{figure}[ht]
\includegraphics[width=10.0cm]{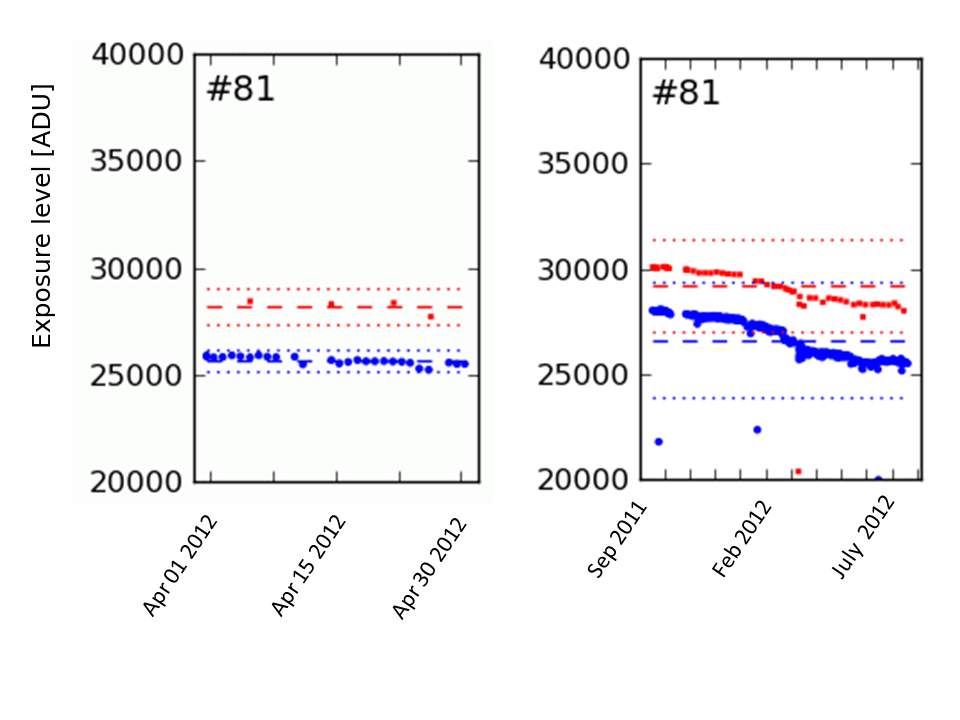}
\caption{Example of trend analysis based on the {\it Quick check} observations for CCDs: \#81. The {\it Quick check} procedure observes the dome screen through the Sloan r filter in a closed dome during daytime. Shown are the raw detector values (dots, in ADU) with their 3$\sigma$ variation (dotted lines) around the clipped mean (dashed line). Both the lamp set in primary use (blue) and secondary use (red) are shown. {\bf Left:} zoom in on April 2012. {\bf Right:} long-term trend for over 10 months of operations.}
\label{f:quickCheck}
\end{figure}

\begin{figure}[ht]
\includegraphics[width=10.0cm]{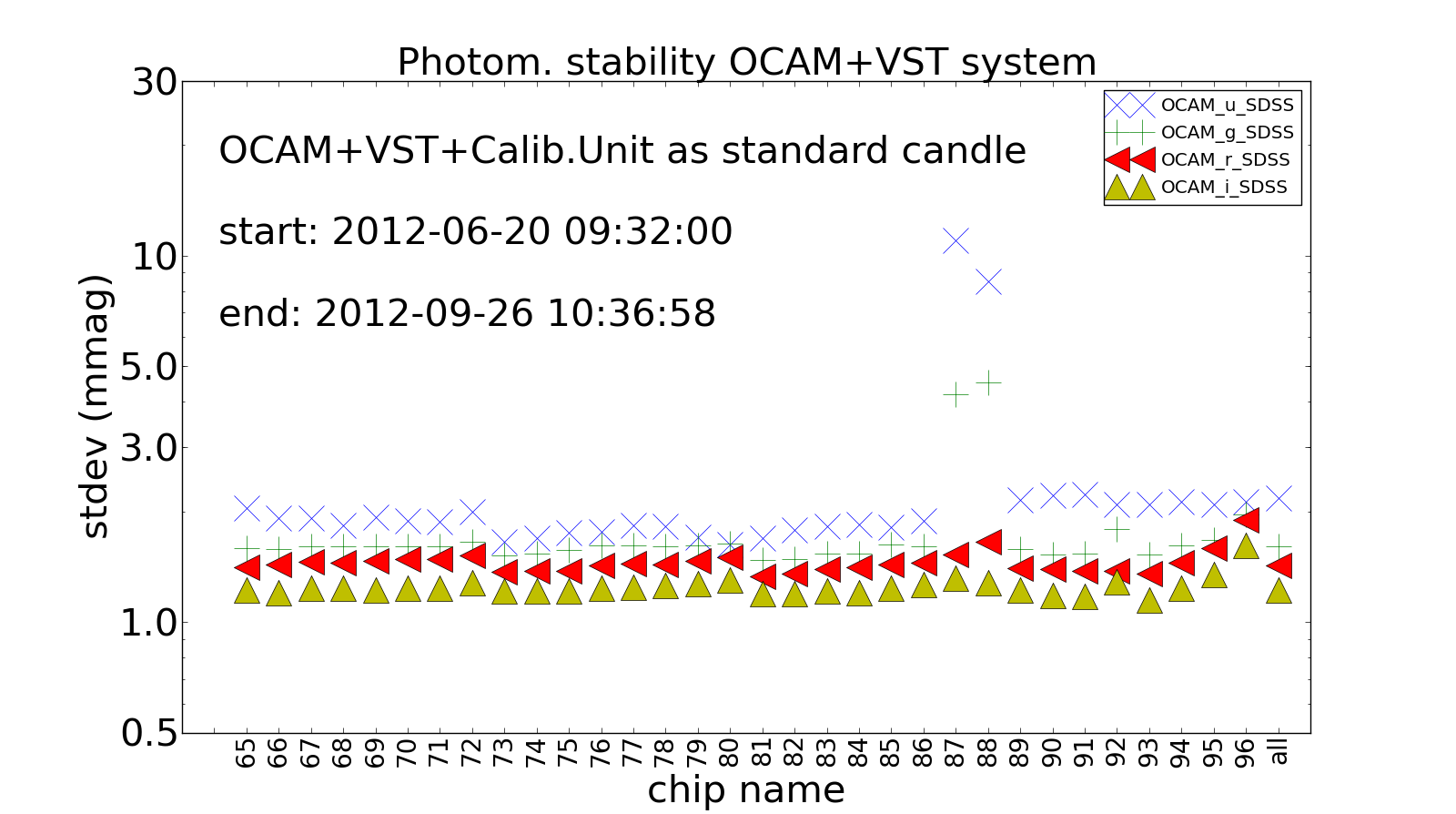}
\caption{Scatter in absolute mean exposure level of domeflats over 3 months for each CCD separately. The large symbols at the end gives the observed scatter for the MOSAIC as a whole. These measurements imply that the in-dome photometric scale is constant to a level of $\sim$1-3mmag level for most CCDs for this period. For the full $\sim$10 months of science operations values have been measured in the range of $\sim$1-6mmag for timescales of a single month.}
\label{f:quickCheckScatter}
\end{figure}

\begin{figure}[ht]
\includegraphics[width=10.0cm]{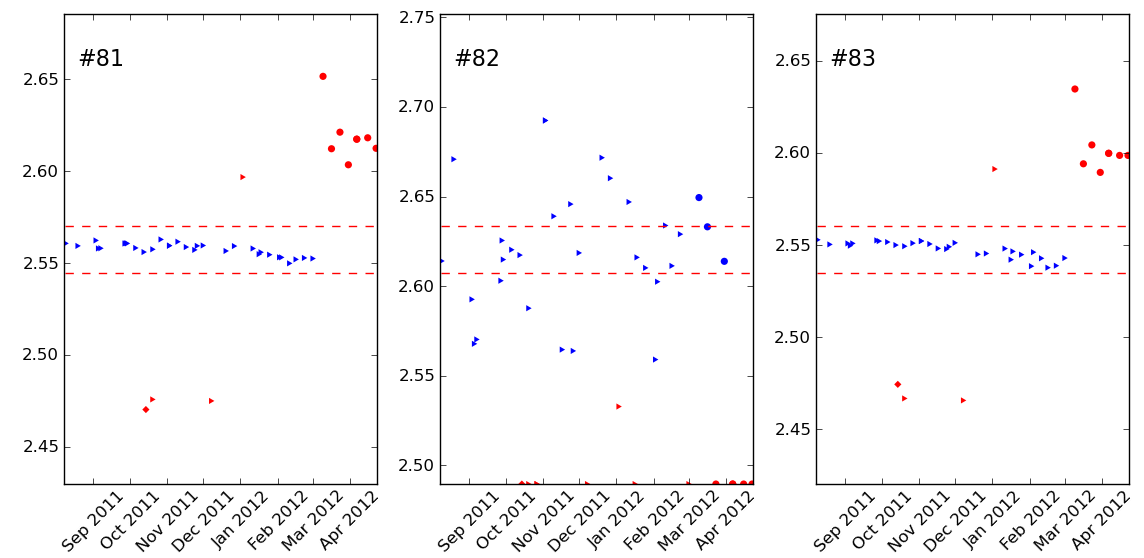}
\caption{Example of a gain monitoring plot for human inspection zooming in on three CCDs: \#81,\#82 and \#83. The measured gain ($e^-$/ADU, dots) is plotted as a function of 8 months of time. Blue/red indicates measurements deemed to be valid/invalid. The red dashed lines indicate $\pm$0.5\% of the median value. The stability observed for \#81 and \#83 is typical for the whole mosaic. The variation in gain measurements for \#82 has disappeared since an intervention on June 2, 2012. The gain procedure uses observations of the dome screen through the Sloan r filter in a closed dome during daytime. They are done at a range of exposure times spanning the full range of electron well depth. The gain is derived from a fit to the Photon Transfer Curve built from those observations.
}
\label{f:gainMonitoring}
\end{figure}

\noindent {\bf Procedure:} The {\it Quick check} provides a daily check on the overall system health in terms of responsivity. It observes the dome screen through the Sloan r filter in a closed dome during daytime for 0.5 seconds. For the operations at Paranal Observatory this measurement leads to the go-ahead/non-conformance verdict in the day report. The {\it Quick check} procedure is a complement to the general domeflats, which are observed weekly for all filters. The {\tt QuickCheckFrame}\footnote{see the Python Documentation server at http://doc.astro-wise.org/astro.main.QuickCheckFrame.html for the description of the {\tt QuickCheckFrame} class.} is the object model implementation of the {\it Quick check} component in the Calibration Plan. It stores data and metadata used in Figure~\ref{f:quickCheck} and provides links to (meta-)data of its dependencies.

\noindent {\bf Usage / preliminary results:} The {\it Quick check} results provide a constraint on the 'in-dome' absolute photometric scale of the system: OmegaCAM, VST and calibration unit. Figure~\ref{f:quickCheck} shows the trend of raw detector levels over one and ten months for one CCD. It implies that the photometric scale of the calibration unit varies almost linear on timescale of a month with a more complex behavior on timescale of several months. The repeatability of the measurement on a monthly scale corresponds to $\sim$ 2-5 mmag per CCD on the system photometric scale for most CCDs. Similarity across chip and filters of residuals to the mean exposure level suggests variability of the lamp is the dominatant source of scatter for most chips. Two CCDs show a larger scatter (see Figure~\ref{f:quickCheckScatter}). These two deviations must originate within electronics specific to these detectors. Previously, CCD\#82 showed a larger variation. This was probably due to gain variations (see Figure~\ref{f:gainMonitoring}). This has been solved with a replacement of a videoboard on June 2, 2012. The origin of the initial decline and subsequent flattening of the responsivity over $\sim$ 10 months is not yet fully understood. Similar behavior has been observed for VLT UT1 and FORS2. A possible reason for this is that the M1 coatings degrade and the M1 becomes dusty (the cause of the primary decline), but
after a few months this reaches a kind of saturation level by which the M1 cannot become much dirtier (ESO private comm.).

\subsection{Dome flat-fielding}
\label{s:domeflatfielding}

\begin{figure}[ht]
\includegraphics[width=7.0cm]{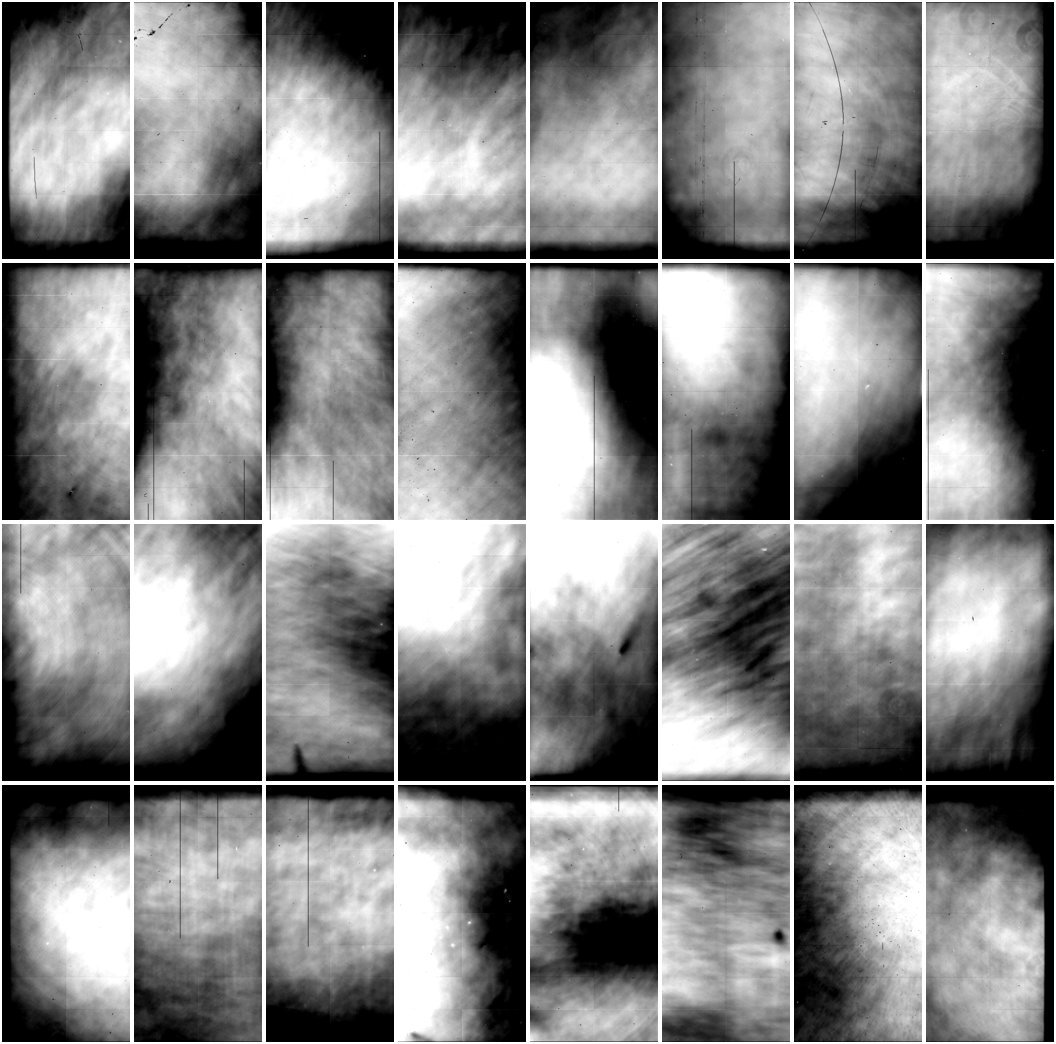}
\includegraphics[width=7.0cm]{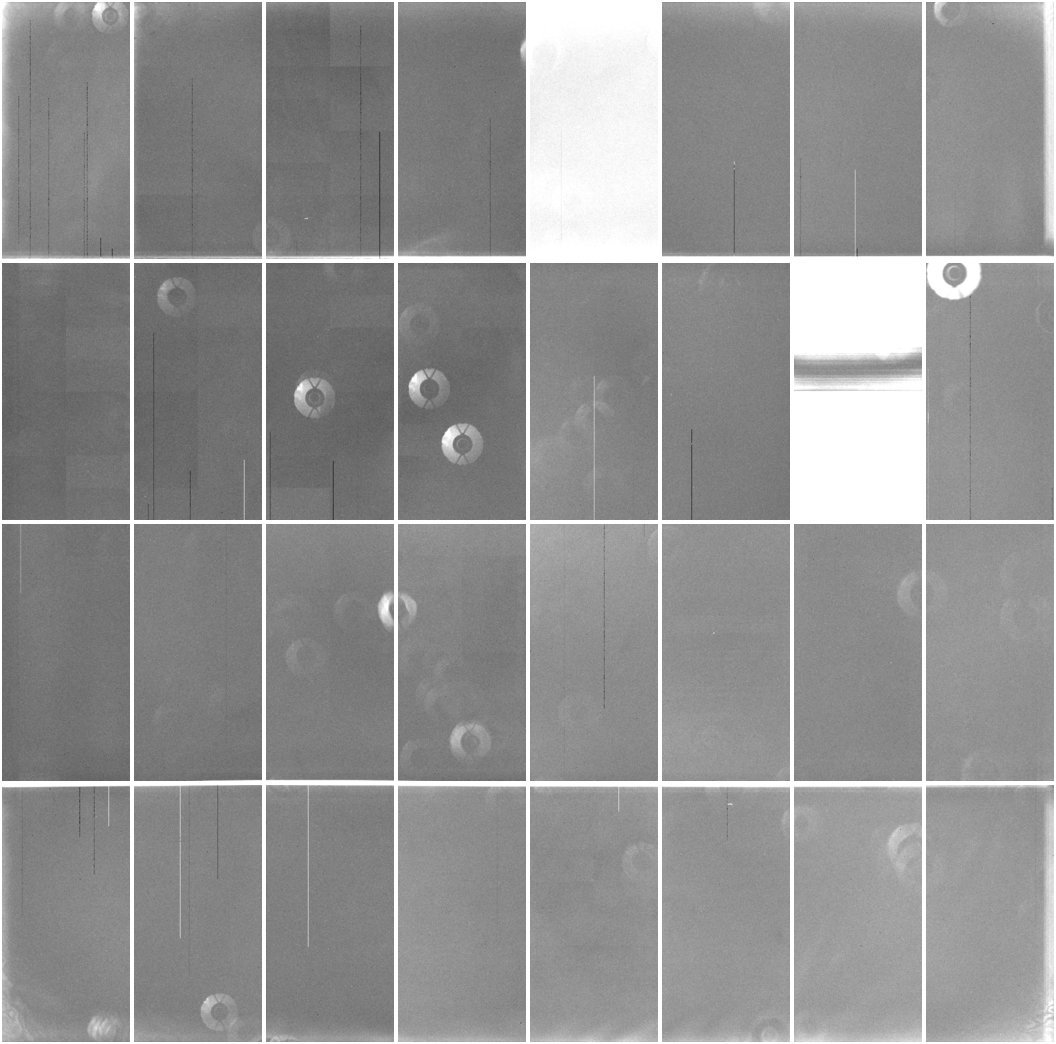}
\caption{{\bf Left:} mean master dome flatfield image through Sloan i from the weekly observations over the period 1 August 2011 - 1 February 2012. Flatfields are per chip normalized to their median. The greyscaling goes from 0.98 (black) to 1.02 (white). {\bf Right:} standard deviation image from the same set of domeflats used left. The grey-scaling goes from 1.0e-3 (black) to 1.6e-3 (white). The typical standard deviation corresponds to $<2$ mmag built-up out of systematic and variable changes in the system excluding the lamp. Two chips (\#82 and \#92) show enhanced variation. Furthermore variations due to changes in bad columns and moving dust can be observed.}
\label{f:domeflat}
\end{figure}

\noindent {\bf Procedure:} Dome flatfields are being measured for the 4 keybands
and the key-composite filter at least once per week. Thus at least within 3 days of the taking
of science data a dome flatfield in the key passband is available. The {\tt DomeFlatFrame}\footnote{see the Python Documentation server at http://doc.astro-wise.org/astro.main.DomeFlatFrame.html for the description of the {\tt DomeFlatFrame} class.} is the object model implementation of the {\it Dome flat-fielding} component in the Calibration Plan.

\noindent {\bf Usage / Preliminary results:} The sequence of master dome flatfields in the keybands, acquired over the months and eventually years is being used to monitor the long term stability of the instrument
and calibration unit (see Figure~\ref{f:domeflat}). From the first 6 months of operations the standard variation in a pixel value of the master domeflat is typically 1.0e-3 - 1.6e-3 in relative units. Changes in bad columns and moving dust are observed to cause standard deviations up to $\sim$3e-3. In conclusion, the variation in master domeflats over 6 months of operations suggest pixel-to-pixel variations in response to correspond to 1-1.5mmag typically on the photometric scale. The variations also put a constraint on the per pixel photometric scale variability of the system without the calibration unit, as the latter one is divided out by normalization. 

\subsection{Twilight flat-fielding}

\begin{figure}[ht]
\includegraphics[width=7.0cm]{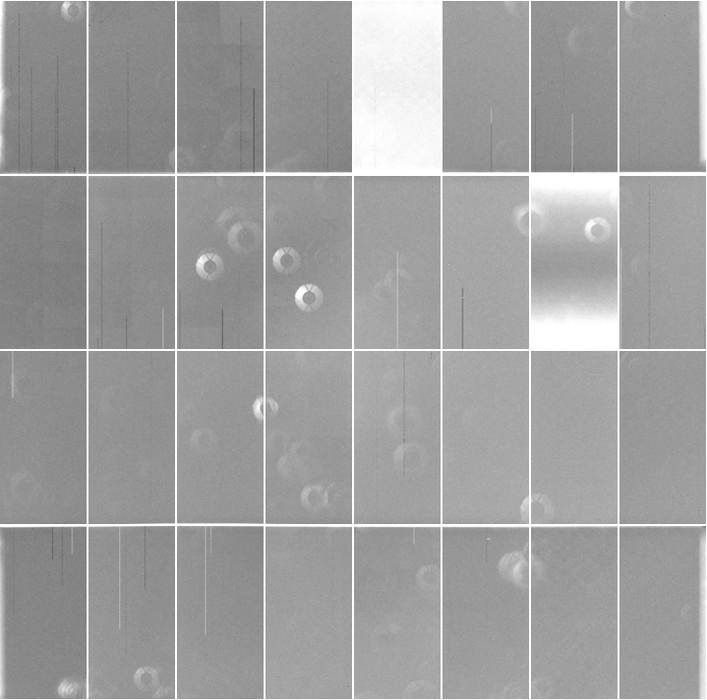}
\includegraphics[width=7.0cm]{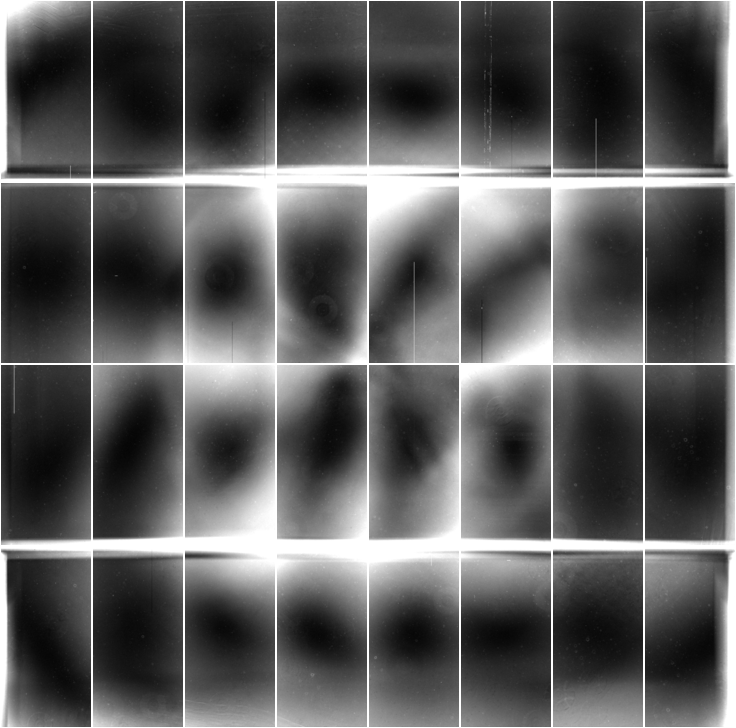}
\caption{{\bf Left:} standard deviation image for the ensemble of master domeflats in Sloan g observed over the period 1 August 2011 - 1 February 2012. The grey-scaling goes from 1.0e-3 (black) to 1.6e-3 (white). The typical standard deviation corresponds to 0.0013 $\pm$ 0.0003 for all CCDs except \#82 and \#92 which have a mean of around 0.0015.  Furthermore variations due to changes in bad columns and moving dust are detectable. {\bf Right:} standard deviation image from the emsemble of twilightflats observed over the same period as the left image. The grey-scaling goes from 2.0e-3 (black) to 6.0e-3 (white). The typical standard deviation is two to three times higher than for domeflats. It is 0.003 $\pm$ 0.001 for the CCDs on the mosaic border and goes up to 0.004 $\pm$ 0.002 for the CCDs in the middle. Variations due to changes in bad columns and moving dust can be observed barely. An additional difference with the domeflats behavior are the higher standard deviations observed near the top and bottom gaps of the three horizontal gaps between the four rows of CCDs. This might be due to scattering and/or vignetting by imperfectly centered masks over the CCD connectors located in these two out of the three gaps (see Figure~\ref{f:omegacamFrontView}).}
\label{f:twilightVsDomeflat}
\end{figure}

\begin{figure}[ht]
\includegraphics[width=10.0cm]{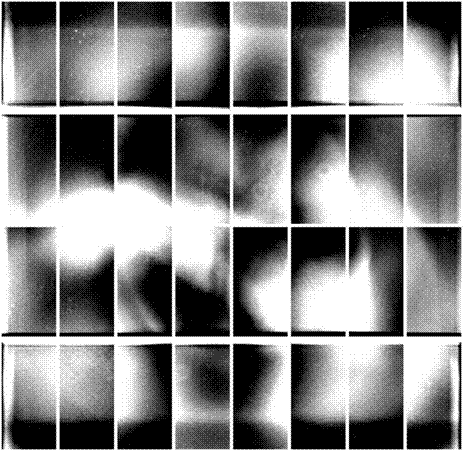}
\caption{Difference image between a single master twilight flat and the average of an ensemble of twilight flats observed at various rotator angles. The greyscaling goes from -0.002 (black) to +0.002 (white). Sequences ordered by rotator angle show systematic behavior of the differences with this angle (see text for reference to movie). This forms the major contribution to the observed standard variations shown in Figure~\ref{f:twilightVsDomeflat}}
\label{f:twilightVsRotatorAngle}
\end{figure}

\noindent {\bf Procedure:} Like domeflats, twilight flatfields are being measured for the 4 keybands
and the key-composite filter at least once per week. Exposure times vary to yield roughly constant exposure levels, i.e., compensating twilight brightness variation. The {\tt TwilightFlatFrame}\footnote{see the Python Documentation server at http://doc.astro-wise.org/astro.main.TwilightFlatFrame.html for the description of the {\tt TwilightFlatFrame} class.} is the object model implementation of the {\it Twilight flat-fielding} component in the Calibration Plan.

\noindent {\bf Usage / preliminary results:} The master twilights acquired over the first $\sim$ 6 months of operations show variations higher by a factor 2 to 3 compared to the domeflats (see Figure~\ref{f:twilightVsDomeflat}). Twilight flats are observed currently at fixed positional angle on the sky and thus at different angles of rotation between instrument and telescope. A significant contribution to the variation observed in flatfields as a function of time is made by systematic changes in the flatfield as a function of rotator angle (see Figure~\ref{f:twilightVsRotatorAngle}). A possible cause is a variation in gradients in the stray-light distribution as a function of angle between OmegaCAM and VST\footnote{See movie of residuals between twilight flatfield at a given rotator angle and the average of an ensemble average at http://wiki.astro-wise.org/projects:omegacam:portal:illuminationcorrection .}. This forms the major contribution to the observed standard variations shown in Figure~\ref{f:twilightVsDomeflat}..

\subsection{Zeropoints}
\label{s:onSky}

\begin{table}
\caption{Standard star fields for zeropoint determination. The number of Landolt \cite{landolt92} and Stetson \cite{stetson00} standard stars within the OmegaCAM FoV are listed. In addition the numbers are listed for available SDSS (DR7) stars and stars from our Preliminary Program (PP) for secondary standards.}
\label{t:saFields}       
\begin{tabular}{|l|r|r|r|r|r|r|}
\hline\noalign{\smallskip}
\noalign{\smallskip}\hline\noalign{\smallskip}
Field & RA [deg] & Dec [deg] & \#Landolt & \#Stetson & \#SDSS & \#PP \\
\hline\noalign{\smallskip}
SA92  &  13.946 &   +0.949   &  41 &   213 &   1094 &  6475  \\
SA95  &  58.500 &   +0.000   &  45 &   426 &   1093 &     0  \\
SA98  & 103.021 &   -0.328   &  46 &  1116 &      0 & 23840  \\
SA101 & 149.112 &   -0.386   &  35 &   117 &   1776 &  5591  \\
SA104 & 190.488 &   -0.529   &  34 &    76 &   1576 &  5701  \\
SA107 & 234.825 &   -0.263   &  28 &   728 &   3889 & 12006  \\
SA110 & 280.600 &   +0.346   &  39 &   589 &      0 & 38562  \\
SA113 & 325.375 &   +0.499   &  42 &   483 &   4046 & 13947  \\
\noalign{\smallskip}\hline
\end{tabular}
\end{table}

\noindent {\bf Procedure:} Zeropoints of the overall detector chain, separately for each CCD chip, in at least all four keybands are obtained nightly irrespective of the science programme. The pointing is selected from a set of 8 equatorial SA fields that cover the full RA cycle (see Table~\ref{t:saFields}). Currently we use as standard catalogs the SDSS DR 8 data in these SA fields and the catalogs from our preparatory programme of Secondary Standards \cite{verdoeskleijn07}. They will be replaced in the future by the OmegaCAM Secondary Standards Catalog on these 8 SA fields. For this, a dedicated OmegaCAM observational programme is in progress. The {\tt PhotometricParameters}\footnote{see the Python Documentation server at http://doc.astro-wise.org/astro.main.PhotometricParameters.html for the description of the {\tt PhotometricParameters} class.} is the object model implementation of the {\it Zeropoints} component in the Calibration Plan. The {\tt PhotRefCatalog}\footnote{see the Python Documentation server at http://doc.astro-wise.org/astro.main.PhotRefCatalog.html for the description of the {\tt PhotRefCatalog} class.} is the object model implementation of the photometric reference catalog. 

\begin{figure}[ht]
\includegraphics[width=7.0cm]{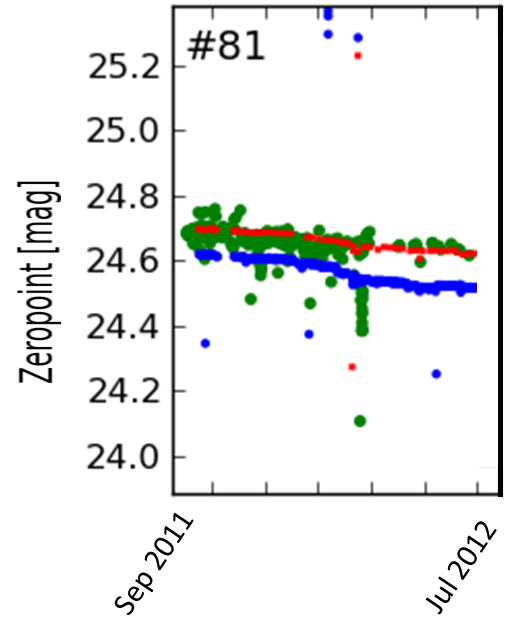}
\caption{Photometric scale in Sloan r on sky and in dome as a function of over 10 months of survey operations for CCD \#81. The green filled circles denote zeropoints obtained from SA field observations. The blue and red filled circles are dome-screen observations from {\tt Quick check} from primary and secondary set respectively. The latter have their average value converted to an arbitrary zeropoint (not a fit!). The similar decrease in zeropoint in time indicated by all three measurements suggests a common cause. See text for more information. The corresponding plot for all 32 CCDs is given AS appendix (Figure~\ref{f:onSkyAndInDomeAll}}
\label{f:onSkyAndInDome}
\end{figure}

\noindent {\bf Usage / preliminary results:} 
We model the photometric behavior of OmegaCAM with the common photometric equation for astronomical imagers:
\begin{equation}
m_{\rm inst} = -2.5\log(countrate) + ZPT - k \times AM 
\label{e:photom}
\end{equation}
where m$_{\rm inst}$ is the magnitude of the object in the instrumental photometric system, the countrate is in ADU/s\footnote{To convert to $e^-$/s: the typical gain for OmegaCAM;s detectors is $\sim$ 2.5-2.6e$^-$/ADU.} , k is the atmospheric extinction coefficient, AM is the airmass and ZPT the zeropoint. We neglect color terms at the moment.
Table~\ref{t:photometricScale} lists the observed scatter in on-sky measurements of the zeropoint, ZPT on timescales of an hour. This is inferred from series of back-to-back SA field observations with chip-sized offsets. This gives an upper-limit on how well we can measure the photometric scale. If the response of VST + OmegaCAM + atmosphere did not vary over the observations, this equals the standard deviation in the photometric scale. Thus it also provides a limit on the contribution of systematic error on the photometric scale. For example, we assume default atmospheric extinction coefficients in this particular case. Further contributions can come from systematics in the reference catalog (SDSS was used here) and the applied illumination correction. The hourly scatter in the on-sky photometric scale per CCD is $\sim$2 -7 times larger than the monthly scatter observed for the in-dome photometric scale.  
A lower-limit to how well we can measure the photometric scale is given by the zeropoint measurement precision. Table~\ref{t:photometricScale} shows that the current precision of the on-sky measurement is 1-3 times larger than the monthly in-dome scatter. The on-sky measurement can be improved by improving the standard catalog (SDSS DR8 catalogs in the SA fields at the moment) and improving the source extraction (aperture magnitudes with fixed 6.4 arcsec diameter, regardless of seeing).
In conclusion, the best ZPT per-chip at any time can be obtained using in-dome observations with the calibration unit. However, this scale is not tied to any photometric system. This might be achieved by scaling the in-dome zeropoint ZPT-D using a set of in-dome and on-sky measurements obtained over many months as constraint. In figures~\ref{f:onSkyAndInDome} and \ref{f:onSkyAndInDomeAll} an illustrative scaling (not a fit!) is overplotted on the on-sky ZPTs. It gives hope that contamination from variability in atmosphere and lamp can be overcome in tying the in-dome photometric scale to a photometric system in the near future.

\begin{table}
\caption{
On-sky and in-dome accuracies and precisions in measuring the zeropoint. ZPT is defined as in Equation~\ref{e:photom}. ZPT-D denotes the in-dome measurement of a zeropoint taking the calibration lamp as a 'standard candle' reference. Stdev(ZPT)$_{\rm MOSAIC}$ is the standard deviation in zeropoints for the 32 CCDs together. Stdev(ZPT)$_{\rm CCD}$ is the same measurement, but now per CCD.  Err(ZPT)$_{\rm CCD}$ is the zeropoint measurement precision derived from the observed scatter of calibrators around the zeropoint.
(The Err(ZPT)$_{\rm MOSAIC}$ is $\sqrt{32}\sim$5.7 smaller). 
}
\label{t:photometricScale}       
\begin{tabular}{|l|r|r|r|r|r|r|}
\hline\noalign{\smallskip}
system & Quantity & timescale & u & g & r & i \\
         &           & &(mmag)&(mmag)&(mmag)&(mmag)\\
\noalign{\smallskip}\hline\noalign{\smallskip}
on-sky & stdev(ZPT)$_{\rm MOSAIC}$ & hour & 11 & 10 & 12 & 11 \\
in-dome& stdev(ZPT-D)$_{\rm MOSAIC}$ & month & 4 & 4 & 3 & 2 \\
\hline
on-sky & stdev(ZPT)$_{\rm CCD}$ & hour & 37 & 10 & 16 & $\sim$15 \\
in-dome & stdev(ZPT-D)$_{\rm CCD}$ & month & 6 & 4 & 3 & 2 \\
\hline
on-sky & Err(ZPT)$_{\rm CCD}$ & - & 15 & 5 & 5 & 7 \\
in-dome & Err(ZPT-D)-$_{\rm CCD}$ & - & $<<1$ & $<<1$ & $<<1$ & $<<1$ \\
\noalign{\smallskip}\hline
\end{tabular}
\end{table}

\subsection{Atmospheric monitoring}

{\bf Procedure:} the OmegaCAM Calibration Plan includes {\tt Atmospheric Monitoring} of  a field near the Equatorial South Pole. Every night, OmegaCAM observes three times the same field near the equatorial South pole: at the beginning, middle and end of a night. This is done through a composite filter containing a Sloan u, g, r and i quadrant. Table~\ref{t:polarField} lists the observational specs for the polar field. Additionally a polar standards photometric catalog is made from dedicated observations. The {\tt PhotometricExtinctionReport}\footnote{see the Python Documentation server at http://doc.astro-wise.org/astro.main.PhotometricExtinctionReport.html for the description of the {\tt PhotometricExtinctionReport} class.} and {\tt PhotSkyBrightness}\footnote{see the Python Documentation server at http://doc.astro-wise.org/astro.main.PhotSkyBrightness.html for the description of the {\tt PhotSkyBrightness} class.} are the object model implementations of the {\it Monitoring} component in the Calibration Plan.

\begin{figure}[ht]
\includegraphics[width=10.0cm]{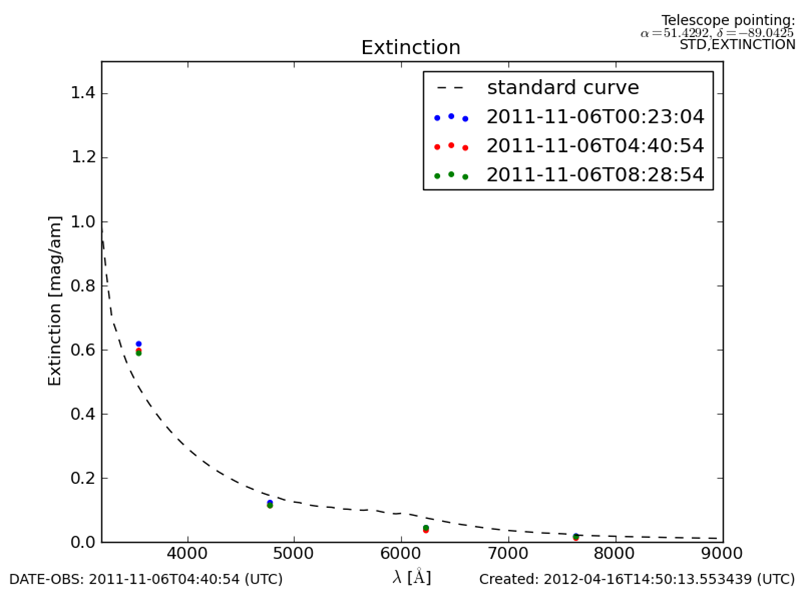}
\caption{Inspection plot for derived excess extinction
(blue points) which is the excess above the standard atmospheric extinction (dashed line).}
\label{f:polarField}
\end{figure}

\begin{table}
\caption{Polar field monitoring specifications}
\label{t:polarField}       
\begin{tabular}{|l|l|}
\hline\noalign{\smallskip}
Topic & Description  \\
\noalign{\smallskip}\hline\noalign{\smallskip}
Coordinates & $\alpha=$51.429167, $\delta$=-89.0425 (J2000) \\
Filter & 4 quadrant glass filter: $u$,$g$,$r$, and $i$ \\
Exposure time & 100s \\
Airmass & 2.3-2.5 \\
\noalign{\smallskip}\hline
\end{tabular}
\end{table}

\noindent {\bf Usage:}
These observations serve two main goals. First, they are used to measure the stability of the nightly atmospheric extinction. This is done by combining information from the {\tt Zeropoints} observations on SA fields and assuming standard atmospheric extinction coefficients. Thus it is possible to determine 3/night the actual atmospheric extinction compared to photometric conditions (Figure~\ref{f:polarField}). Second, the observations are used to measure skybrightness. Again these can be compared to expected values under photometric conditions for given lunar phase/distance. Long-term monitoring of both the atmospheric extinction results and sky brightness results can yield deeper understanding in the physics setting the atmospheric extinction (dust, aerosols, water vapor etcetera). Eventually, this should lead to survey calibrations without the need to assume default atmospheric extinction coefficients.

As the maximum difference in airmass is 8\% between any two polar field observations it directly measures changes in the atmospheric extinction coefficient at fixed high airmass to this accuracy. Similarly, comparison of overlapping sources in dithered science observations measure changes in atmospheric extinction coefficients at fixed lower airmasses. 
Results from Sections~\ref{s:inDome} and \ref{s:onSky} show that the zeropoint ZPT could be constrained independently by combining information from {\tt Quick health} and {\tt Zeropoints} observations. The combination of information from SA fields ({\tt Zeropoints}) at typical airmass$\sim 1.2$ and polar field ({\tt Atmospheric Monitoring}) at airmass$\sim 2.4$ can then constrain the behavior of the atmospheric extinction coefficient k as a function of time from a nice baseline in airmass.

\subsection{Illumination characterization}
\begin{figure}[!ht]
\includegraphics[width=6.4cm,angle=-90]{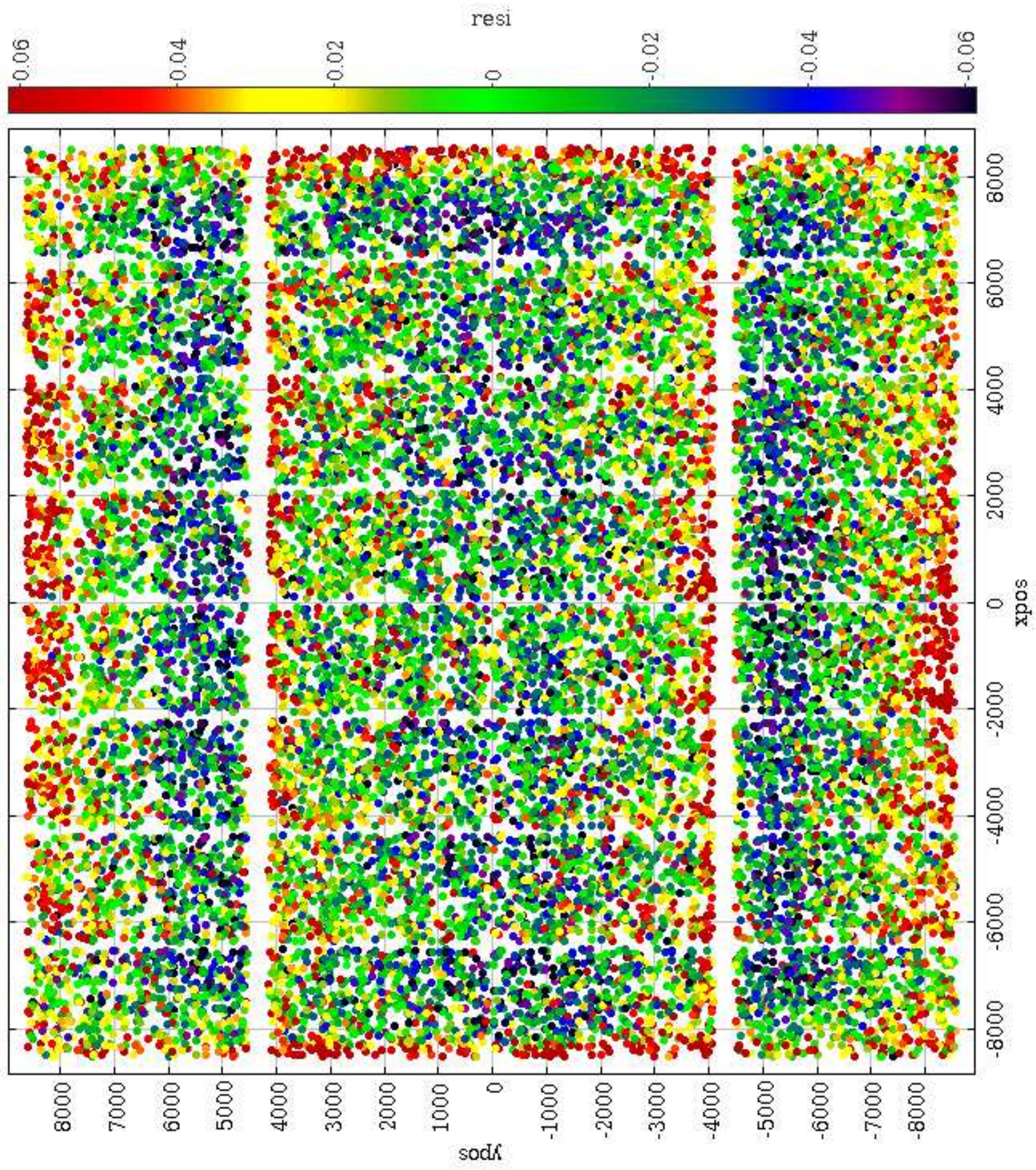}
\includegraphics[width=6.4cm,angle=-90]{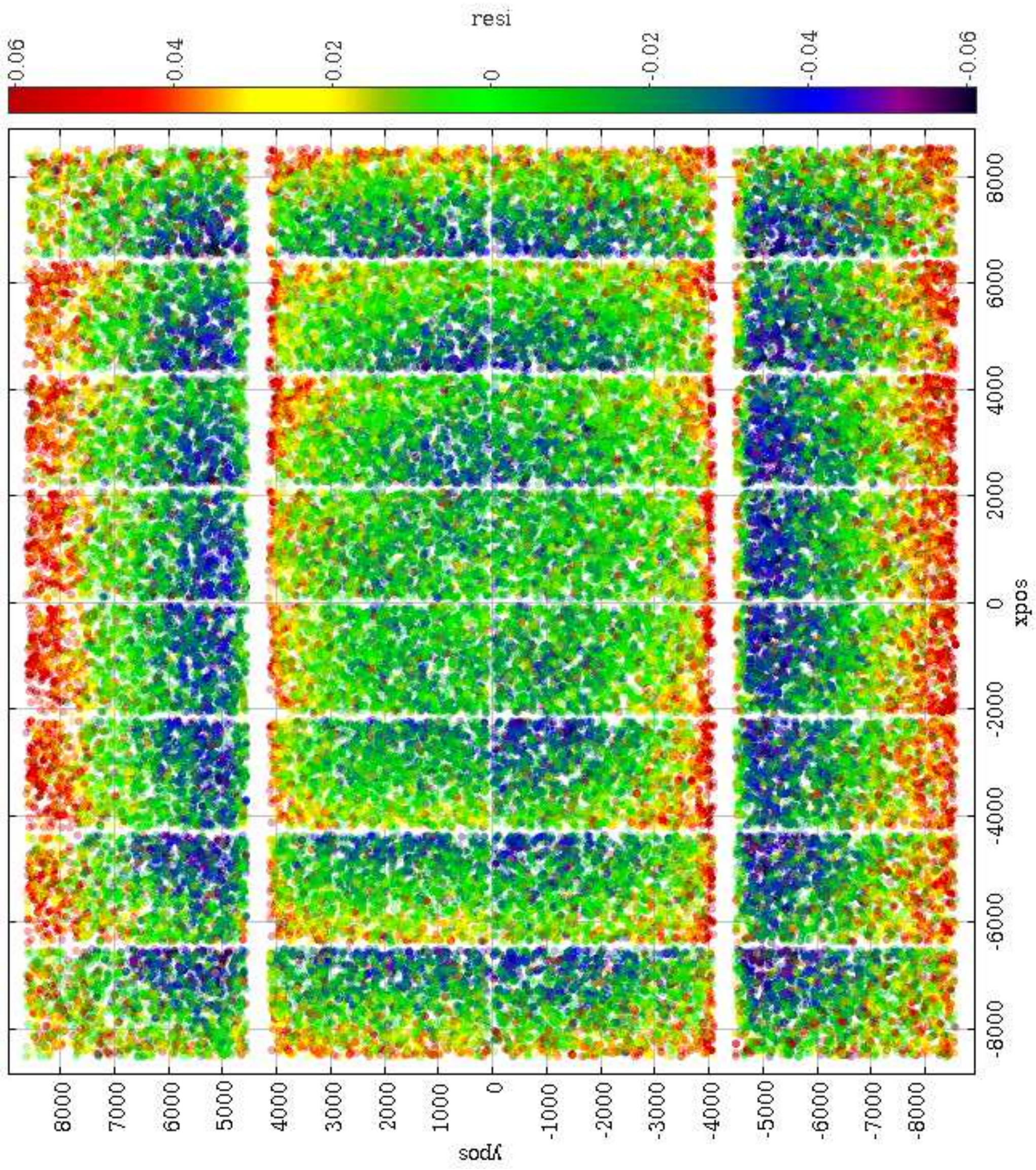}
\includegraphics[width=6.4cm,angle=-90]{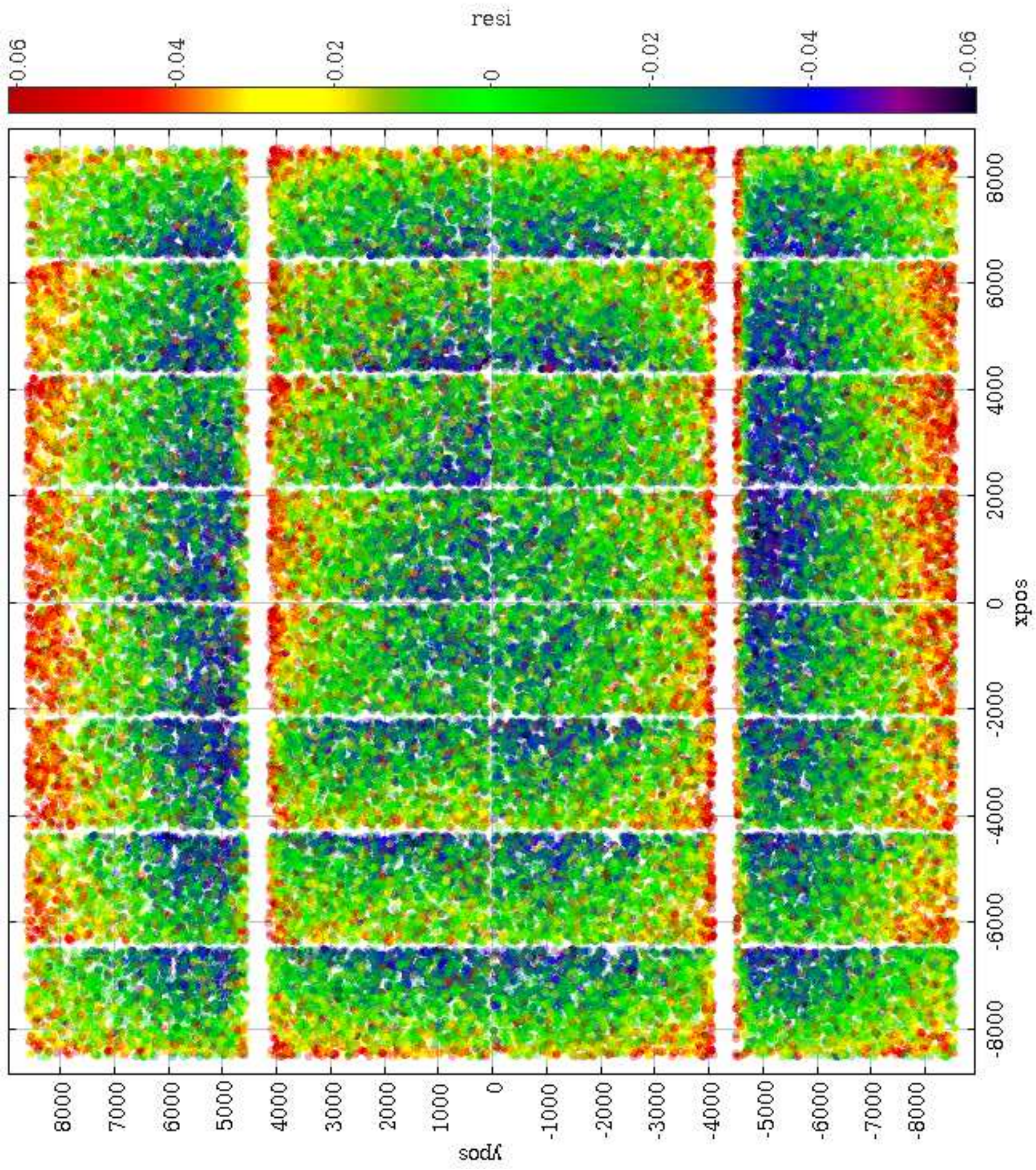}
\includegraphics[width=6.4cm,angle=-90]{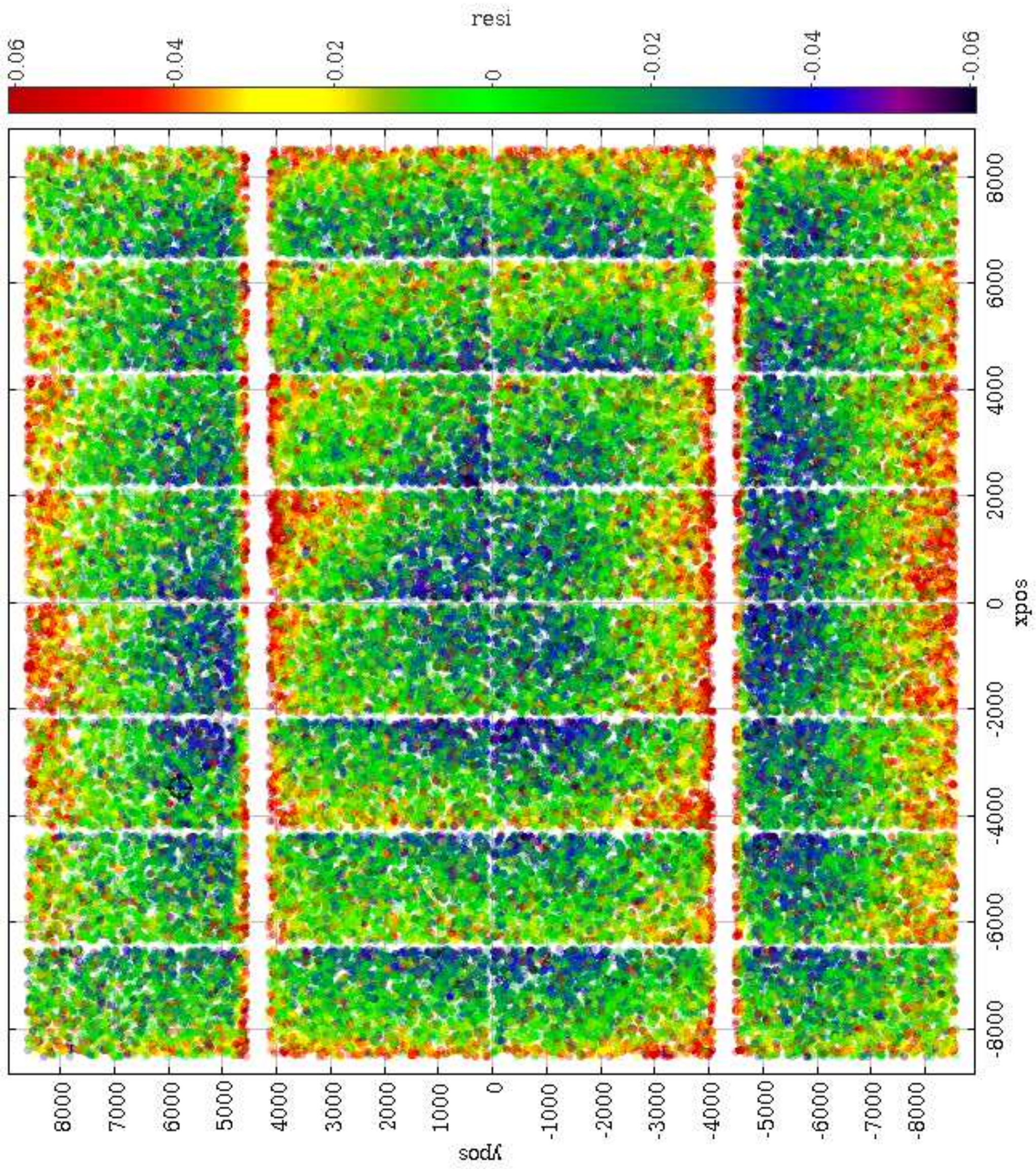}
\includegraphics[width=6.4cm,angle=-90]{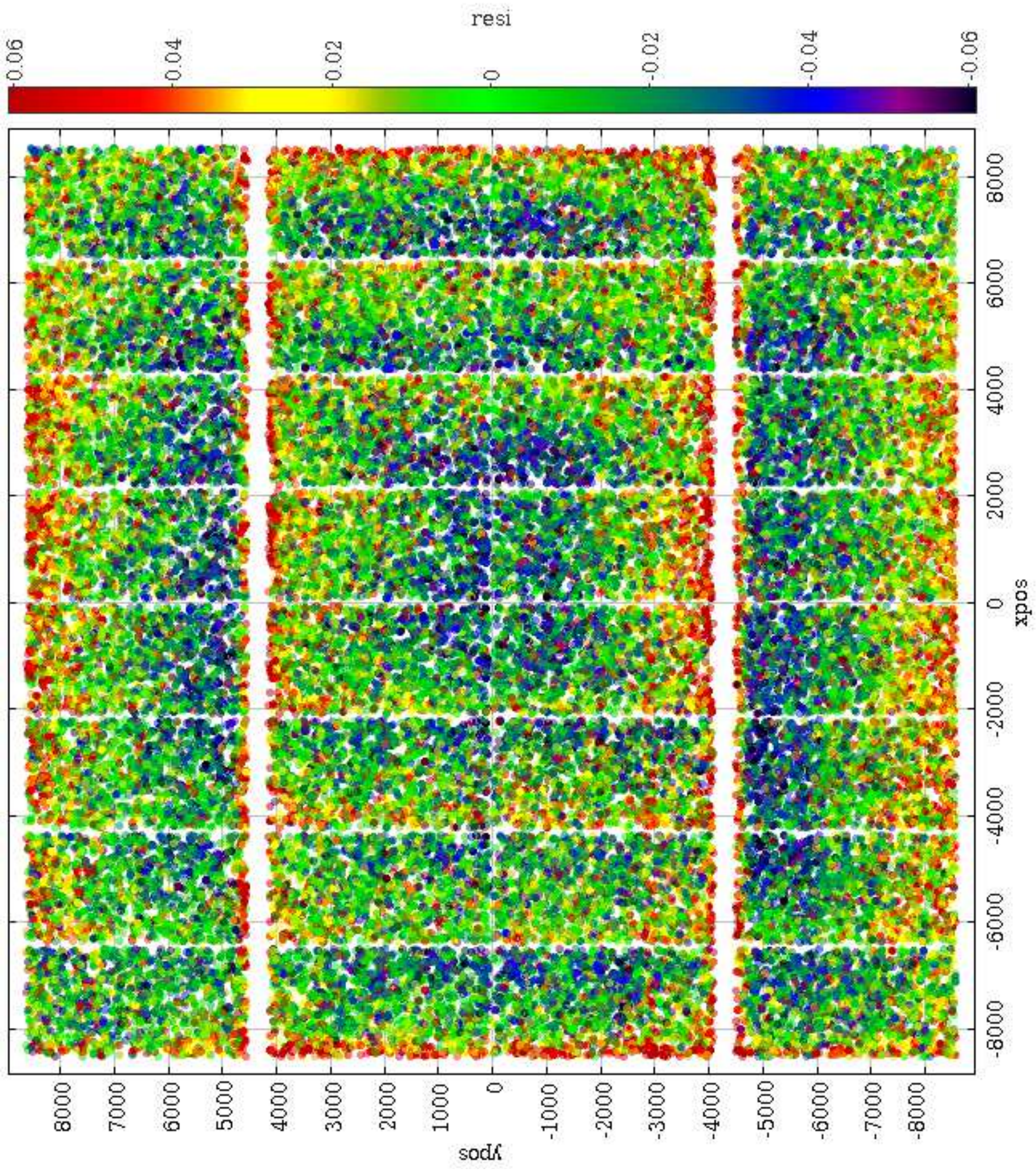}
\caption{Photometric residual of stars in SA113 after subtraction of an individual zeropoint per chip. Top-left to bottom-right: Sloan u,g,r,i and z. Residuals are computed using  magnitudes for stars listed in the SDSS DR7 catalog. Residuals are in AB magnitudes in the OmegaCAM instrumental system. No color correction has been applied yet to convert the SDSS DR7 catalog to this system.}
\label{f:illumVariations}
\end{figure}

Commissioning analysis has shown that twilight and dome flat exposures with OmegaCAM are both subjected to non-uniform illumination. Thus, the derivation of a 'true' flatfield (i.e., only the variation in pixel sensitivity) requires a correction of these illumination variations. This is termed illumination correction\footnote{A detailed description of the illumination characterization and correction is given in commissioning document VST-TRE-OCM-23100-3608 (available at http://wiki.astro-wise.org/projects:omegacam:portal:illuminationcorrection)}. Here we give a summary. If no correction for these illumination variations is applied the photometric scale residual can vary by up to $\sim 0.1$ mag, bottom-to-peak within a single chip (Figure~\ref{f:illumVariations}). The large-scale residual pattern appears roughly point symmetric around the center of the mosaic. A variation in the stray light distribution which is point symmetric around the optical axis to first order can cause this. Other instruments show similar illumination variations patterns, i.e., having a central-axisymmetric 2D-polynomial shape, for example, WFI \cite{manfroid01} and~\cite{koch04} and  MEGACAM~\cite{regnault07}. In addition there are localized components, especially near the covers over the detector heads. Finally, commissioning analysis has shown a dependency on the rotator angle between the camera and the telescope. From trend analysis on $\sim$5 months of weekly dome flats and twilight flats the conclusion is that the intrinsic flat field (i.e., pixel sensitivities) varies with a  $\sigma \sim 0.2\%$ or less. Therefore it was decided to build a single masterflat from a combination of a single twilight flat and single dome flat for the Sloan g, r, i and z filters and apply this to all OmegaCAM observations used to determine the illumination correction\footnote{For u only the twilight flat is used see commissioning document VST-TRE-OCM-23100-3608 (available at http://wiki.astro-wise.org/projects:omegacam:portal:illuminationcorrection) for an explanation}. The usage of a single flat (i.e., at single rotator angle) simplifies the procedure: a single illumination correction needs to be derived.

\begin{figure}
\includegraphics[width=6.7cm,angle=-90]{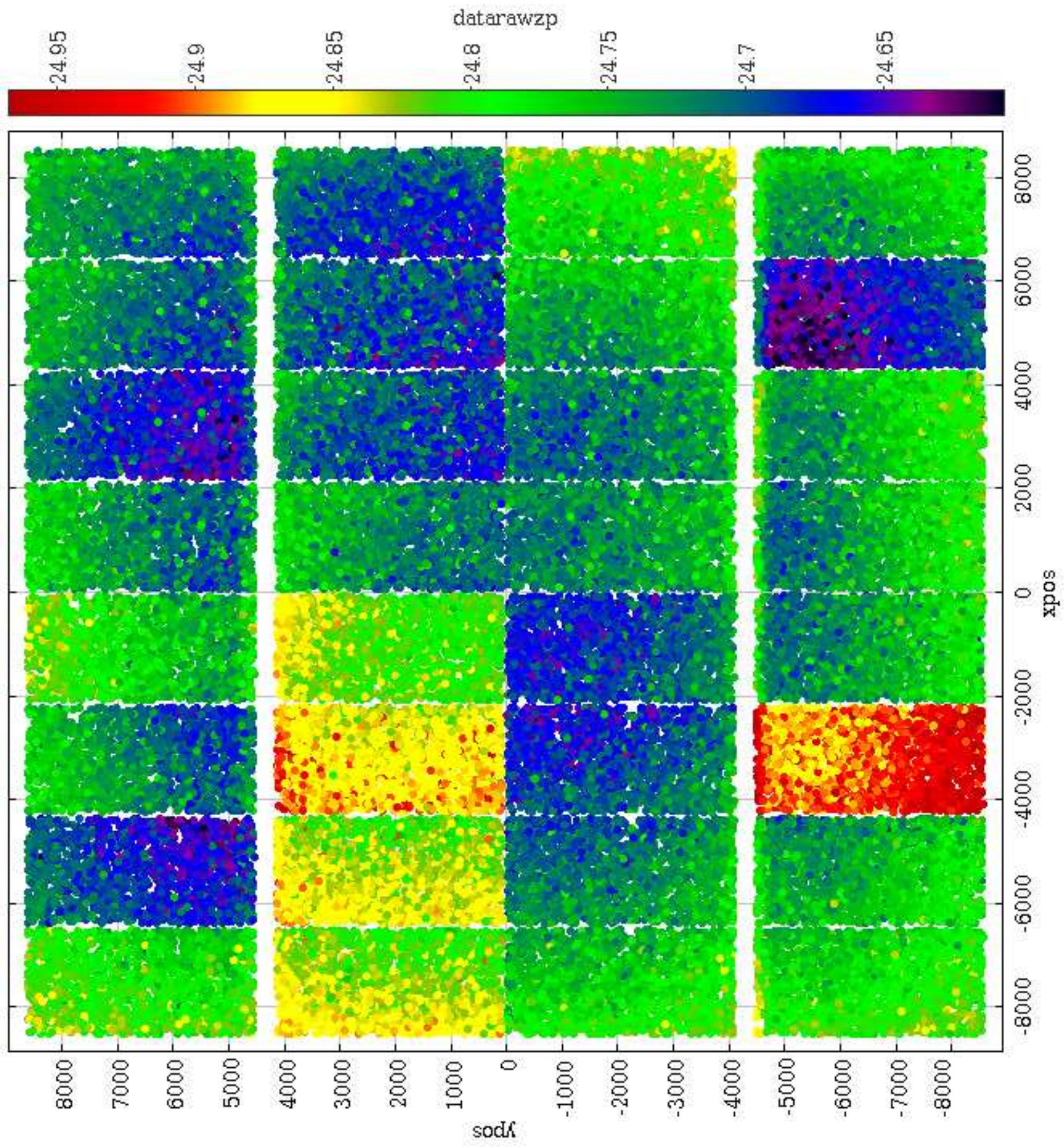}
\includegraphics[width=6.7cm,angle=-90]{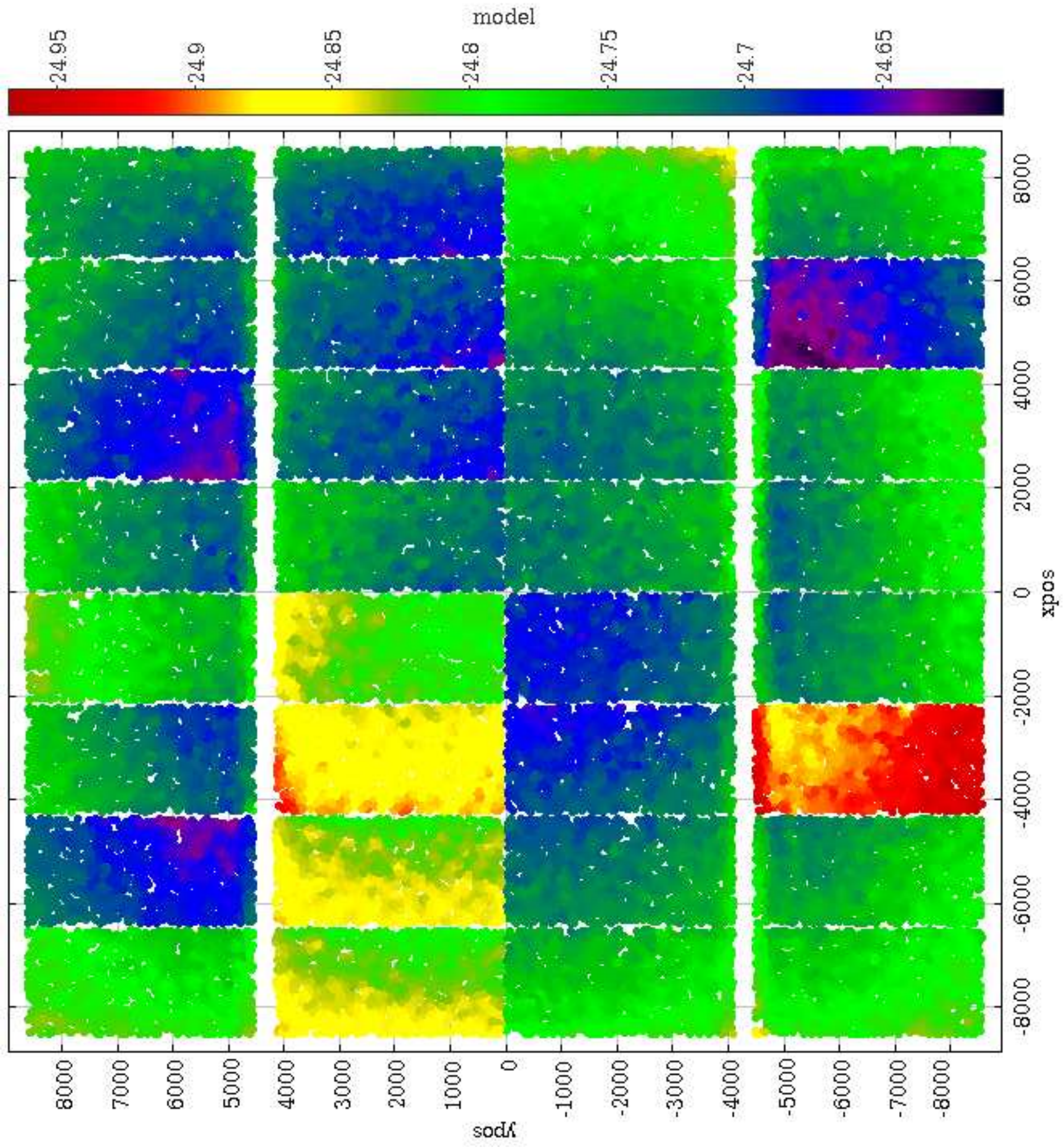}
\includegraphics[width=6.7cm,angle=-90]{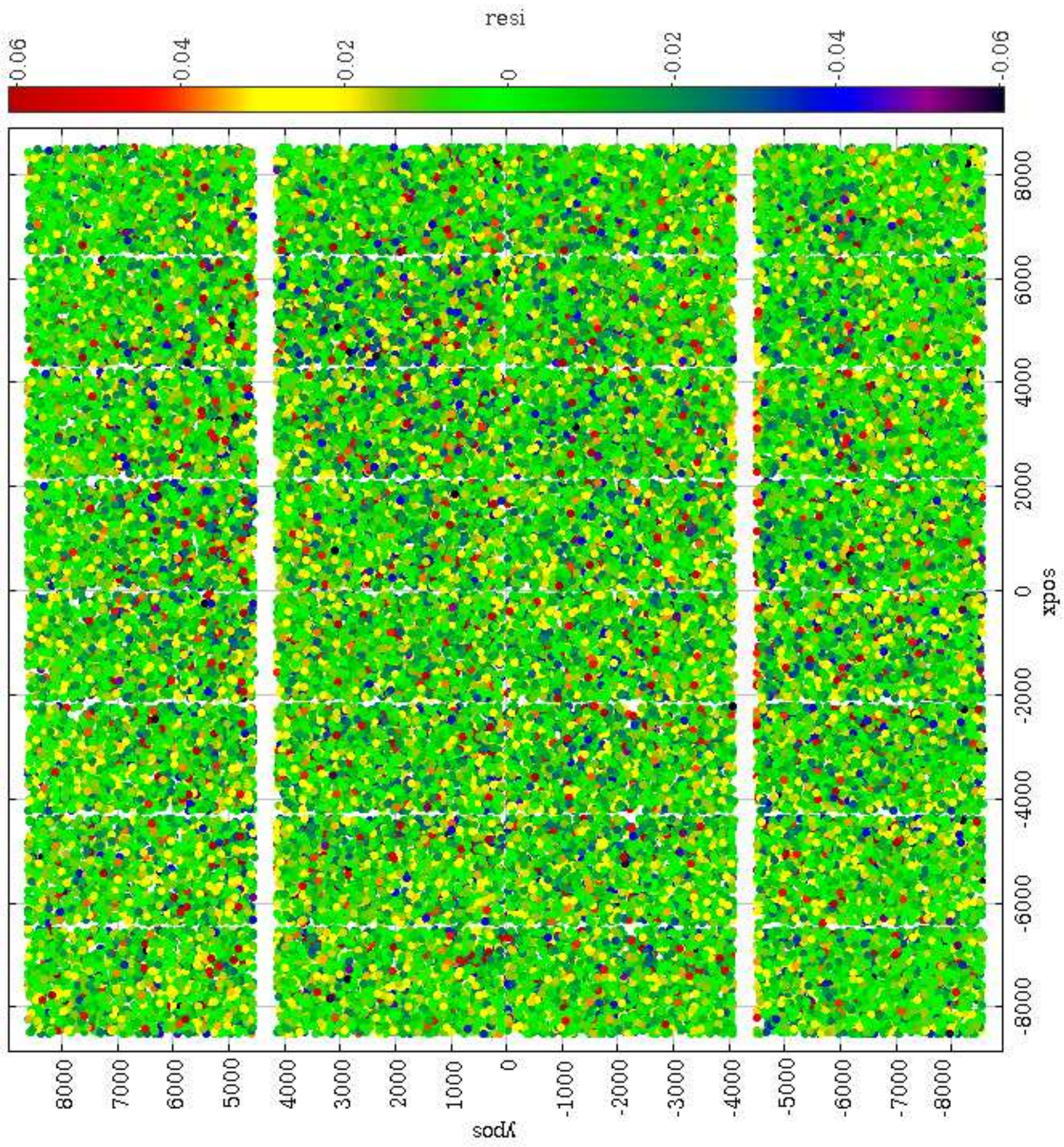}
\includegraphics[width=6.7cm,angle=-90]{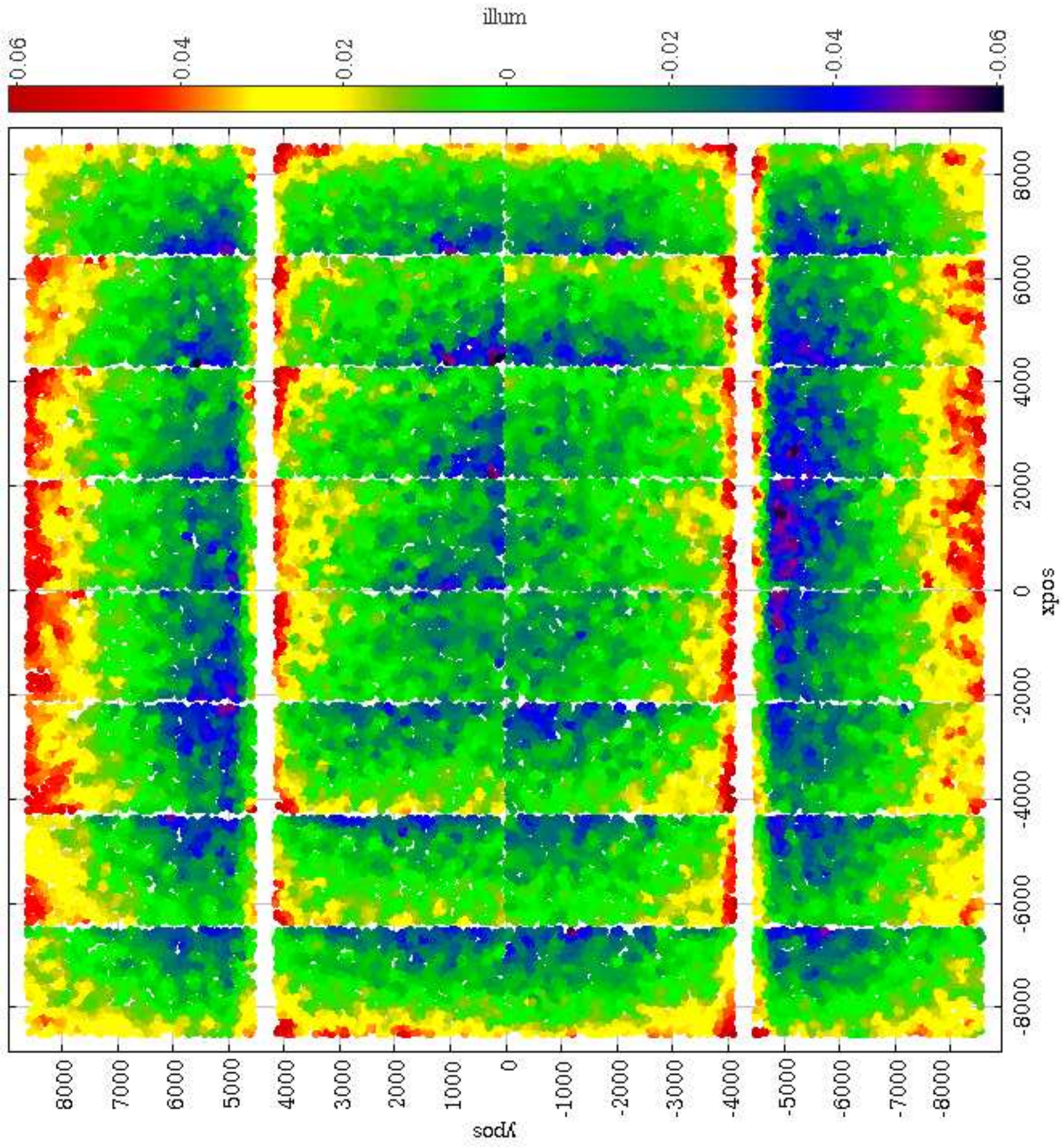}
\includegraphics[width=6.7cm]{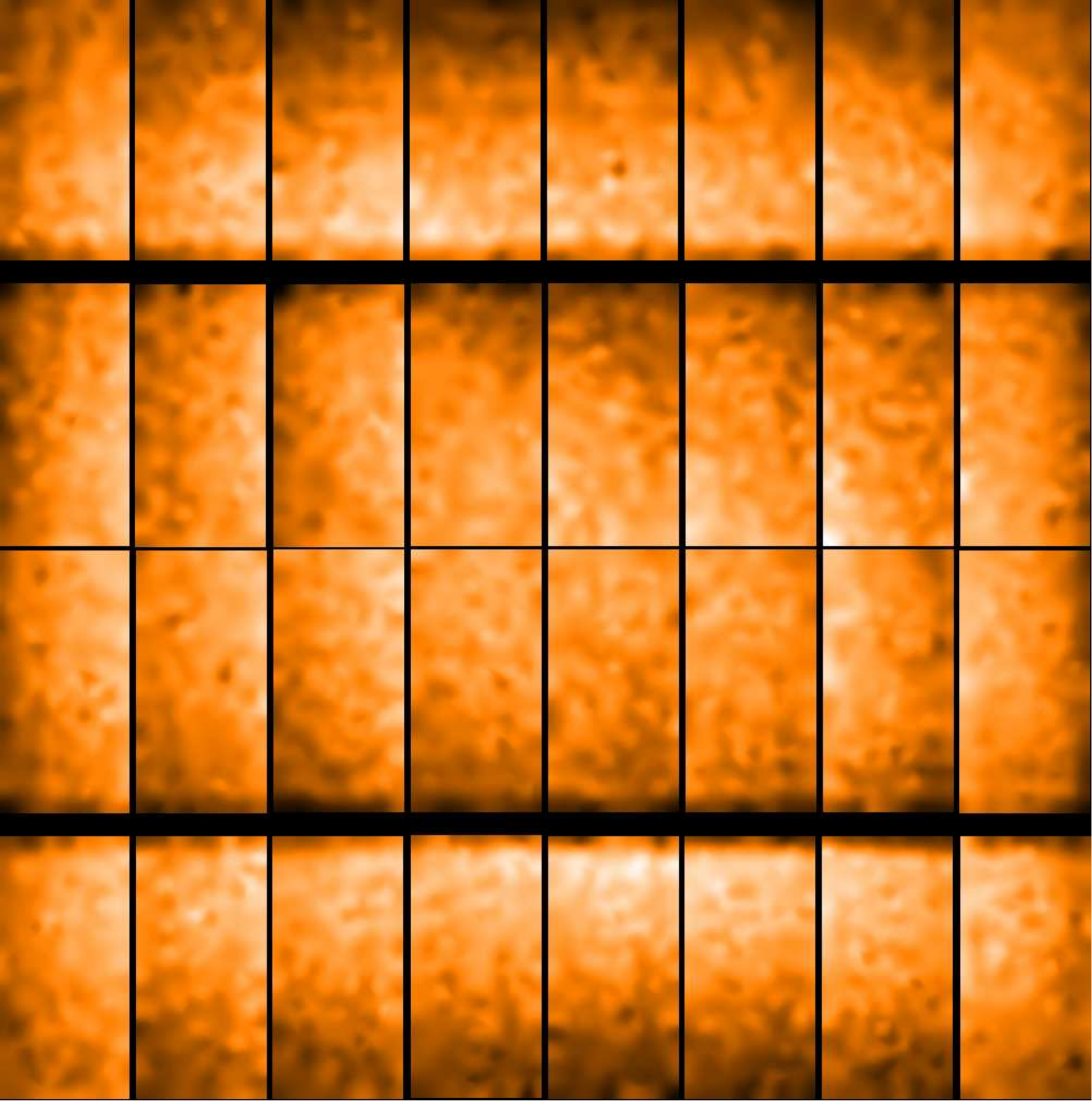}
\caption{Illumination correction results for Sloan r projected on the focal plane. The datapoints in the top four plots represent all DR7 reference stars of SA92, SA95 and SA113 after dithering 33 times and filtering (see text). Top left: raw zeropoints for all SDSS DR7 reference stars. Top right: model = fitted zeropoints per chip + illumination variation model. Middle left: residual magnitudes of reference stars after applying illumination correction. Middle right: illumination variation model (i.e. with ZPT per chip subtracted, similar to Figure~\ref{f:illumVariations}). Bottom: the resulting illumination correction pixelframe as applied in the photometric data processing.}
\label{f:illuminationCorrection}
\end{figure}

To characterize the illumination variations SA fields were observed at $\sim$ 33 dither positions. These dither positions are mostly single-chip offsets such that a group of stars is observed on all 32 CCDs. Zeropoints per chip are computed and the magnitude residuals, i.e., the difference between the zeropoint and reference magnitude, sample the illumination variations. The combined SA fields, yield roughly 1000 useful magnitude residuals per chip for g, r, i ,z and about 500 for the u-band. The residuals are modelled as a continuous locally linear model using adaptive binning and interpolation. The results are summarized in Table~\ref{t:illuminationCorrection} and shown in Figure~\ref{f:illuminationCorrection}.

\begin{table}[!ht]
\begin{minipage}{\textwidth}
\begin{center}
\caption{{\bf Illumination Correction Results}. Standard deviations of residual magnitudes in the five SLOAN filters left after applying the localized illumination corrections. The internal and external standard deviation (sd) of the magnitude residuals are defined as follows: external is with respect to the stellar reference magnitude in the standard star catalog (SDSS DR7 stellar photometry). Internal is w.r.t. the mean of its observed magnitudes. 2nd and 3rd column: sd of internal and external errors if only zero points are applied; 4th and 5th columns: sd of internal and external errors when illumination correction applied on OmegaCAM KiDS survey data (SA fields for z band).}
\label{t:illuminationCorrection}
\begin{tabular}{|r||r|r|r|r|}
\hline
      &  zp only& zp only& illum.corr.  &  illum.corr. \\
\hline
(1) & (2)& (3)& (4) & (5) \\
\hline
band  &sd\_int & sd\_ext & sd\_int & sd\_ext \\
\hline
   &(mmag) & (mmag) & (mmag) & (mmag) \\
\hline
u & 29 & 36 & 13& 25 \\
g & 20 & 30 &  9& 24 \\
r & 21 & 29 & 12& 22 \\
i & 22 & 30 & 10& 24 \\
z & 27 & 33 & 15& 27  \\
\hline
\end{tabular}
\end{center}
\end{minipage}
\end{table}

The illumination correction is characterized to better than 1\% for the amplitude over a single CCD. The correction method overcomes the rotator angle dependence by virtue of the stability of the intrinsic (``true'') flatfield. This illumination correction has remained accurate on timescales of at least 7 months of OmegaCAM survey operations (including a VST technical intervention). The illumination correction approach here can account for the sub-percent changes in pixel-to-pixel sensitivity over time. This can be done by updating the domeflat because it sets the small-scale structure in the masterflat. 

The observed flatfield and the illumination correction derived for it form a multiplicative pair. Together they describe the true pixel sensitivity over the field. The multiplication procedure does not remove the additive component of straylight itself. In other words, the correction should be applied to sources that are small compared to the scale over which variation in straylight strength is significant. Sources with more extended surface brightness distributions on scales of a CCD or larger in science exposures are still contaminated by straylight. In this case a combination of a global illumination correction plus local background fitting appears able to remove the straylight contamination \cite{grado12}. If no such dedicated correction is applied the extended surface brightness distributions are affected systematically up to 0.04mag for typical cases. 

Similar to flatfielding and zeropoints the illumination correction is represented in the Astro-WISE data model. The {\tt IlluminationCorrection}\footnote{see the Python Documentation server at http://doc.astro-wise.org/astro.main.IlluminationCorrection.html for the description of the {\tt IlluminationCorrection} class.} is the object model implementation of the {\it IlluminationCorrection} component in the Calibration Plan. 

\section{Conclusions and Outlook}
\label{s:conclusionsAndOutlook}

OmegaCAM at the VST has been commissioned successfully, meeting specifications and started science operations October 15, 2011. The continuous flow of observations from OmegaCAM's Photometric Calibration Plan are being processed in Astro-WISE. Astro-WISE is also used for photometric monitoring and analyses. The first half a year of photometric calibrations yield the following preliminary results:
\begin{itemize}
\item The on-sky measurements with OmegaCAM achieve 10mmag precision in sampling the photometric scale of OmegaCAM+VST+atmosphere.
\item The in-dome photometric scale of detector + calibration unit is stable over a month to within $\sim$5mmag for $\sim$90\% of the detectors. 
\item Master domeflat variations over 6 months of operations constrain the relative pixel-to-pixel responses to be constant to a level of 1-1.5mmag.
\item Measuring the VST+OmegaCAM response with in-dome observations (domeflats) can be done with $\sim$2-5 times higher precision than with on-sky observations (standard fields). First results give hope that by combining the results from in-dome and on-sky observations we can model variations in the atmospheric extinctions coefficients very accurately.
\end{itemize}

The Astro-WISE system is also being used also to handle the Kilo Degree Survey. Observing conditions are partially non-photometric. The goal of our OmegaCAM photometric calibration is to reach photometric homogeneity to 1\% accuracy per filter and between filters. KiDS observes 1500 square degrees of sky in two patches. The area is tiled with $\sim$1 square degree observations that have some overlap. The tiling is identical for each filter. These tiles are constructed per filter from an $\sim 1$ hour observing block of dithered observations, 4 in u and 5 in g, r and i. This survey design yields constraints on variations in photometric scale in several ways. Variations in the photometric scale on the hour scale are sampled by dither-to-dither comparisons within an observing block. The overlaps between KiDS tiles provide intra-night and inter-night constraints. The ATLAS Public Survey running in parallel on OmegaCAM at the VST covers KiDS survey area in the same filters to a shallower depth, using a different tiling. The overlaps within KiDS and between KiDS and ATLAS provide therefore additional mostly inter-night constraints. Exploiting these constraints to improve survey photometric homogenization is an on-going effort. 

It is the combination of the total volume of photometric information over years of calibrations of OmegaCAM and survey observations of KiDS and ATLAS that can lead to an internally consistent continuous description of the observatory's photometric scale to the best accuracy. The Astro-WISE system design is aimed specifically at combining such large volumes of complex information dynamically, improving the solution as science and calibration observations progress and new information accumulates in Astro-WISE.

\begin{acknowledgements}
The OmegaCAM consortium was formed in response to an announcement of opportunity from ESO, and comprises
institutes in the Netherlands (NOVA, in particular the Kapteyn Institute Groningen and Leiden Observatory),
Germany (in particular University Observatories of Munich, Gottingen and Bonn) and Italy (INAF, in particular
Padua and Naples observatories). The ESO Optical Detector Team designed and built the detector system.
OmegaCAM is headed by PI K. Kuijken (Leiden University).
We acknowledge the good support by ESO staff during OmegaCAM commissioning, with special thanks to Steffen Mieske and Dietrich Baade.
OmegaCAM is funded by grants from the Dutch Organization for Research in Astronomy (NOVA), the German
Federal Ministry of Education, Science, Research and Technology (grants 05 AV9MG1/7, AV9WM2/5, 05
AV2MGA/6 and 05 AV2WM1/2), and the Italian Consorzio Nazionale per l'Astronomia e l'Astrofisica (CNAA)
and Istituto Nazionale di Astrofisica (INAF), in addition to manpower and materials provided by the partner
institutes. OmegaCAM dataflow operations in The Netherlands are supported by Target (www.rug.nl/target).

\end{acknowledgements}

\begin{appendix}

\begin{figure}[ht]
\includegraphics[width=20.0cm, angle=-90]{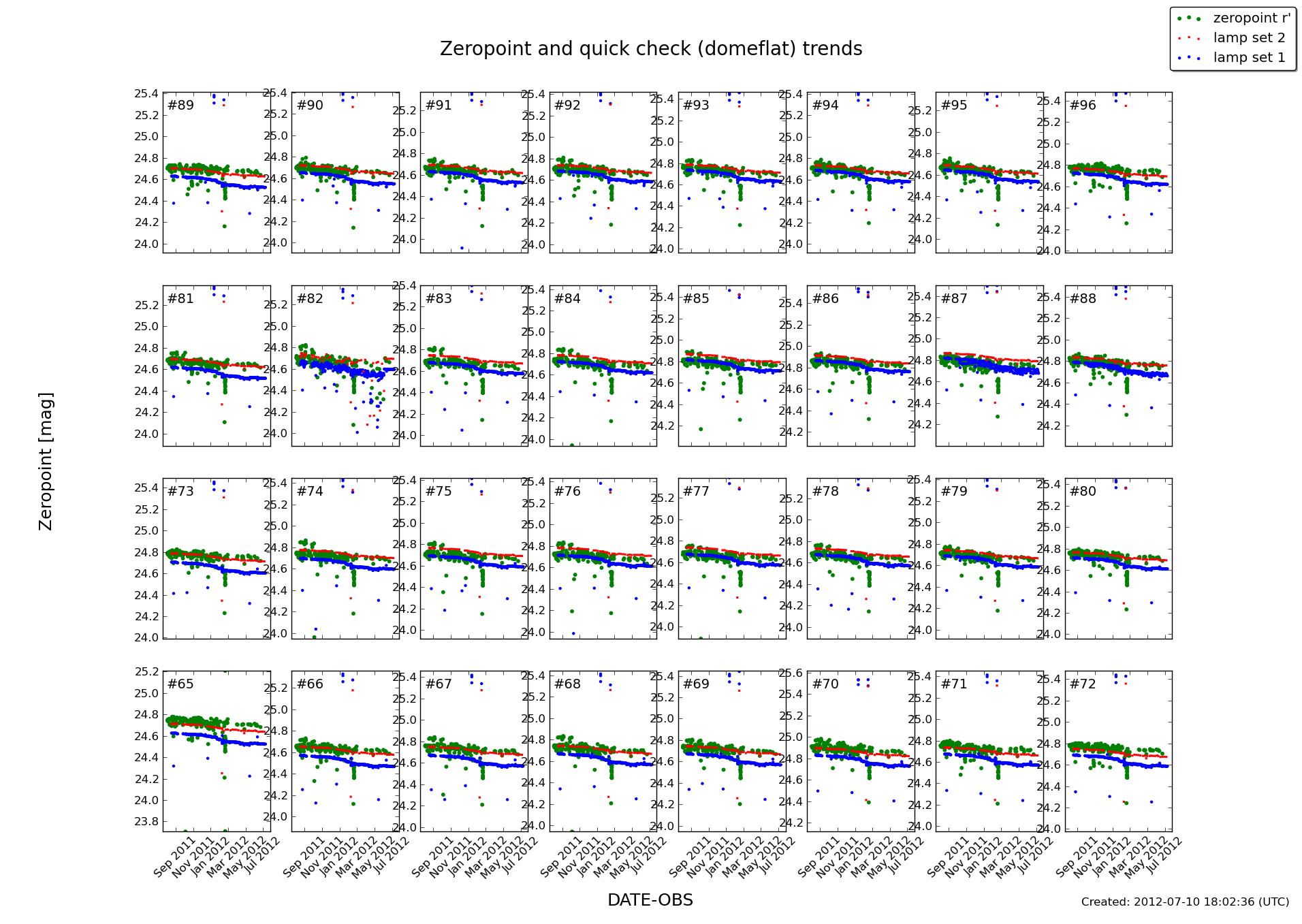}
\caption{Photometric scale in Sloan r on sky and in dome as a function of over 10 months of survey operations for all 32 science CCDs. The green filled circles denote zeropoints obtained from SA field observations. The blue and red filled circles are dome-screen observations from {\tt Quick check} from primary and secondary set respectively.The latter have their average value converted to an arbitrary zeropoint (not a fit!). See Section~\ref{s:onSky} for a discussion.}
\label{f:onSkyAndInDomeAll}
\end{figure}

\end{appendix}




\end{document}